\documentclass[aps,prd,twocolumn,superscriptaddress,showpacs]{revtex4}
\usepackage{epsfig}
\usepackage{amssymb}
\usepackage{graphicx}
\usepackage{dcolumn}
\newcommand{\beq}{\begin{equation}}
\newcommand{\eeq}{\end{equation}}
\newcommand{\beqar}{\begin{eqnarray}}
\newcommand{\eeqar}{\end{eqnarray}}
\newcommand{\ds}{\displaystyle}
\usepackage{color}

\begin{document}


\title{How many of the scaling trends in $pp$ collisions will be
       violated at $\sqrt{s_{\rm NN}} = 14$\,TeV ? - Predictions 
       from Monte Carlo quark-gluon string model.}

\author{J.~Bleibel}
\affiliation{
Institut f\"ur Angewandte Physik, Universit\"at T\"ubingen,
Auf der Morgenstelle 10, D-72076 T\"ubingen, Germany
\vspace*{1ex}}
\affiliation{
Max-Planck-Institut f\"ur Intelligente Systeme, 
Heisenbergstr. 3, D-70569 Stuttgart, Germany
\vspace*{1ex}}
\author{L.V.~Bravina}
\affiliation{
Department of Physics, University of Oslo, PB 1048 Blindern,
N-0316 Oslo, Norway
\vspace*{1ex}}
\affiliation{
National Research Nuclear University "MEPhI" (Moscow Engineering Physics
Institute), Kashirskoe highway 31, Moscow, RU-115409, Russia
\vspace*{1ex}}
\affiliation{
Frankfurt Institute for Advanced Studies, Ruth-Moufang-Str. 1,
D-60438 Frankfurt a.M., Germany
\vspace*{1ex}}
\author{E.E.~Zabrodin}
\affiliation{
Skobeltzyn Institute for Nuclear Physics,
Moscow State University, RU-119899 Moscow, Russia
\vspace*{1ex}}
\affiliation{
Department of Physics, University of Oslo, PB 1048 Blindern,
N-0316 Oslo, Norway
\vspace*{1ex}}
\affiliation{
National Research Nuclear University "MEPhI" (Moscow Engineering Physics
Institute), Kashirskoe highway 31, Moscow, RU-115409, Russia
\vspace*{1ex}}
\affiliation{
Frankfurt Institute for Advanced Studies, Ruth-Moufang-Str. 1,
D-60438 Frankfurt a.M., Germany
\vspace*{1ex}}

\date{\today}

\begin{abstract}
Multiplicity, rapidity and transverse momentum distributions of hadrons
produced both in inelastic and nondiffractive $pp$ collisions at 
energies from $\sqrt{s} = 200$\,GeV to 14\,TeV are studied within the 
Monte Carlo quark-gluon string model (QGSM). Good agreement with the 
available experimental data up to $\sqrt{s} = 13$\,TeV is obtained, and 
predictions are made for the collisions at top LHC energy 
$\sqrt{s} = 14$\,TeV. The model indicates that Feynman scaling and 
extended longitudinal scaling remain valid in the fragmentation regions, 
whereas strong violation of Feynman scaling is observed at midrapidity. 
The Koba-Nielsen-Olesen (KNO) scaling in multiplicity 
distributions is violated at LHC also. The origin of both maintenance
and violation of the scaling trends is traced to short range 
correlations of particles in the strings and interplay between the 
multistring processes at ultrarelativistic energies.
\end{abstract}


\maketitle

\section{Introduction}
\label{sec1}

The recent interest in general features of elementary hadronic 
interactions, especially in characteristics of
$pp$ collisions, at ultrarelativistic energies is manifold. First of
all, these collisions are conventionally used as reference ones to 
reveal the nuclear matter effects, such as strangeness enhancement,
nuclear shadowing, collective flow, etc., attributed to formation of a 
pattern of hot and dense nuclear matter and the quark-gluon plasma 
(QGP) in the course of ultrarelativistic heavy-ion collisions (see
\cite{QM14} and references therein). Although the formation of the 
QGP and/or collective behavior was not found yet in $pp$ collisions
at energies up to the Tevatron energy $\sqrt{s} = 1.8$\,TeV, 
strong evidence for azimuthal correlations up to $\sqrt{s} = 
7$\,TeV has been reported \cite{Abelev:2013sqa}, and physicists are 
discussing the possibility to observe, e.g., elliptic flow in $pp$ 
interactions at $\sqrt{s} = 7$\,TeV and $\sqrt{s} = 13$\,TeV 
accessible for the Large Hadron Collider (LHC) at CERN at present. This 
limit may be raised to $\sqrt{s} = 14$\,TeV in the nearest future. 
Because of the huge amount of energy deposited in the overlapping 
region, the $pp$ systems might be similar to $A+A$ collisions at nonzero 
impact parameter at lower energies \cite{CDP08,Ent10} and, therefore, 
demonstrate collectivity. An alternative approach developed in 
\cite{BKK08} considers the flow effects in hadronic interactions as 
initial state effects linked to correlation between the transverse 
momentum and position in the transverse plane of a parton in a 
hadron. 
In recent paper \cite{GLMT16} the authors argue that 
elliptic flow in $pp$ collisions stems from the density variation 
mechanism within the Color Glass Condensate (CGC) saturation physics.
This important problem should definitely be clarified in the future. 

Then, the problem of multiparticle production in elementary hadronic 
collisions is not fully solved yet. Here, for hard processes with large
momentum transfer, the running coupling constant $\alpha_S$ is small, 
and that allows for application of the perturbative quantum chromodynamics
(QCD). For soft processes with small momentum transfer, which give 
dominant contribution to high energy hadronic interactions, the
$\alpha_S$ is close to unity and therefore, nonperturbative methods 
should be applied. Many microscopic models
\cite{fritiof,pythia,mc_qgsm,dpm,venus,hijing,urqmd,epos,phojet,qgsjet2,
paciae}
based on the string picture of particle production \cite{jetset}
have been successfully employed to describe gross features of hadronic
collisions at relativistic and ultrarelativistic energies, whereas the
statistical approach pioneered more than 50 years ago by Fermi and 
Landau \cite{Fer50,Land53} is not ruled out. To make predictions for 
the LHC in the latter case, one has to extrapolate the data obtained at 
lower energies to the high energy region. It was found quite long ago
that, despite the complexity of a reaction with tens or more particles 
in a final state, multiparticle production in $pp$ collisions exhibits
several universal trends, such as $(\ln{\sqrt{s}})^2$ dependence of
total charged particle multiplicity \cite{PDG}, Feynman scaling 
\cite{Feyn_scal} and related to it extended longitudinal scaling 
\cite{ext_long_scal}, Koba-Nielsen-Olesen (KNO) scaling \cite{KNO_scal}, 
and so forth. Similar trends were found later on in proton-nucleus and 
nucleus-nucleus collisions as well (for review see, e.g., 
\cite{Busza08}). On the other hand, the description of ultrarelativistic 
hadronic interactions in the framework of color glass condensate theory
\cite{CGC} leads to a universal power-law behavior of, e.g., density
of produced charged particles per unit of rapidity and their 
transverse momentum \cite{Lev10,MLP10}.

The aim of the present article is to study the main characteristics of
$pp$ interactions at energies from $\sqrt{s} = 200$\,GeV to top LHC
energy $\sqrt{s} = 14$\,TeV. We employ the Monte Carlo (MC) 
realization \cite{mc_qgsm} of the quark-gluon string model (QGSM) 
\cite{qgsm_1} based on Gribov's Reggeon field theory (RFT) 
\cite{RFT} that obeys both analyticity and unitarity requirements. The 
features of the model are described in Sec.~\ref{sec2} in detail. 
Comparisons with available experimental data for $\bar{p}p$ and $pp$ 
collisions at energies $\sqrt{s} \geq 200$\,GeV, including the
measurements at LHC for $pp$ interactions at $\sqrt{s} = 2.36$~TeV, 
$\sqrt{s} = 7$~TeV, and the recently measured $\sqrt{s} = 13$\,TeV, 
as well as predictions for the top LHC energy $\sqrt{s}  = 14$\,TeV, are 
presented in Sec.~\ref{sec3}. Here, exclusive contributions of soft and 
hard processes to particle rapidity and transverse momentum spectra are 
studied. Special attention is given to the origin of violation of the 
KNO scaling, violation of the Feynman scaling at midrapidity, and its 
maintenance in the fragmentation region. Obtained QGSM results are also 
confronted to the predictions of other microscopic and macroscopic 
models. Conclusions are drawn in Sec.~\ref{sec4}.    
 
\section{Quark-gluon string model and its Monte Carlo realization}
\label{sec2}

As was mentioned in the Introduction, the description of soft hadronic 
processes cannot be done within the perturbative QCD. Therefore, 
similarly to the dual parton model \cite{dpm}, 
the quark-gluon string model \cite{qgsm_1} employs the so-called 
$1/N$ series expansion \cite{tH74,Ven74} of the amplitude for 
processes in QCD, where $N$ is either number of colors $N_c$ 
\cite{tH74} or number of flavors $N_f$ \cite{Ven74}. In this approach 
the amplitude of a hadronic process is represented as a sum over 
diagrams of various topologies, so the method is often called {\it 
topological expansion\/}. It appears that at high energies and small 
momentum transfer the arising diagrams are similar \cite{CMV75,Kaid99} 
to processes describing the exchange of Regge singularities in the
$t$-channel. For instance, planar diagrams correspond to the exchange
of Reggeons, and cylinder diagrams correspond to reactions without
quantum number exchange in the $t$-channel, i.e., taking place via the 
Pomeron exchange, where Pomeron is a composite state of the Reggeized
gluons. Processes with many-Pomeron or many-Reggeon exchanges are also
possible. To find the amplitude of multiparticle production, one has to
cut the diagrams in the $s$-channel, and the physical picture of
quark-gluon strings arises. Namely, new particles are produced through 
the formation and breakup of quark-gluon strings or exited objects
consisting of quarks, diquarks and their antistates connected by a 
gluon string.

\begin{figure}[htb]
 \resizebox{\linewidth}{!}{
\includegraphics{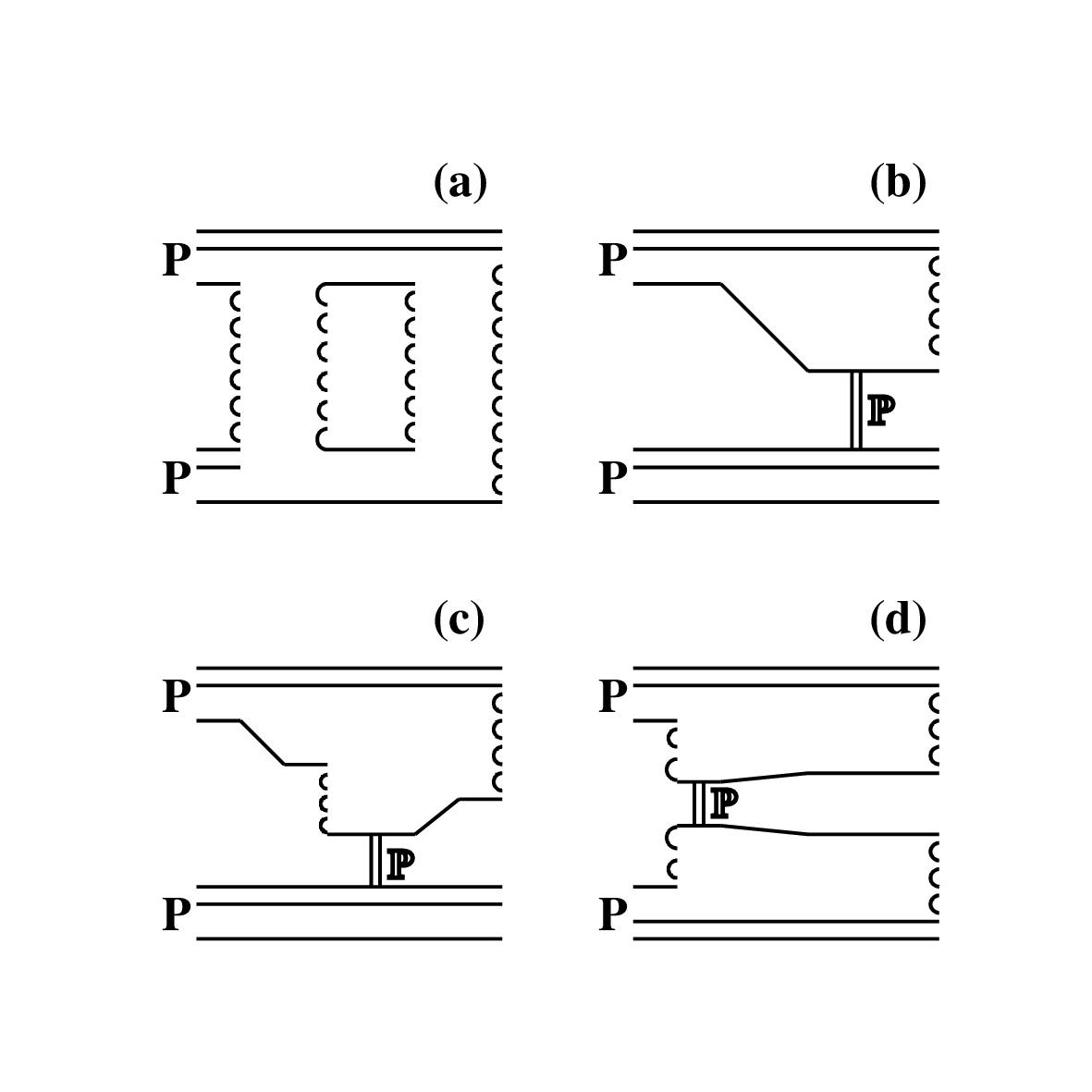}
}
\caption{
Diagrams of particle production processes included in the modeling of 
$pp$ interactions at ultrarelativistic energies. See text for details.}
\label{fig1}
\end{figure}
Figure~\ref{fig1} shows the subprocesses with particle creation taken
into account in the current Monte Carlo version of the QGSM 
\cite{mc_qgsm} for $pp$ collisions at ultrarelativistic energies. The 
inelastic cross section consists of three terms
\beq
\ds
\sigma_{inel}^{pp}(s) = 
                    \sigma_P(s) + \sigma_{SD}(s) + \sigma_{DD}(s)\ ,
\label{eq1}
\eeq
where $\sigma_P(s)$ is the cross section for the multi-chain processes
described by the cylinder diagram and diagrams with multi-Pomeron 
scattering [Fig.~\ref{fig1}(a)], $\sigma_{SD}(s)$ is the cross section 
of single-diffractive processes represented by the diagrams with small 
[Fig.~\ref{fig1}(b)] and large [Fig.~\ref{fig1}(c)] mass excitation, 
corresponding to the triple-Reggeon and triple-Pomeron limit, 
respectively, and $\sigma_{DD}(s)$ is the cross section of 
double-diffractive process shown by the diagram in Fig.~\ref{fig1}(d). 
Other diagrams that are relevant at low and intermediate energies, such 
as the undeveloped cylinder diagram or
diagram with quark rearrangement \cite{mc_qgsm}, play a minor role 
here because their cross sections rapidly drop with rising $s$. 
The statistical weight of each subprocess is expressed in terms of
the interaction cross section for the given subprocess $\sigma_i(s)$,
\beq
\ds
\omega_i = \sigma_i (s) / \sigma_{inel} (s)\ .
\label{eq2}
\eeq
Then, the hadron inelastic interaction cross section 
$\sigma_{inel}(s) = \sigma_{tot}(s) - \sigma_{el}(s)$ is split into
the cross section for single-diffractive interactions $\sigma_{SD}(s)$
and the cross section for nondiffractive reactions $\sigma_{ND}(s)$,
as is usually done in analysis of experimental data. 
By means of the Abramovskii-Gribov-Kancheli (AGK) cutting rules 
\cite{AGK} the inelastic nondiffractive interaction cross section 
$\sigma_{ND}(s)$ can be expressed via the sum of the cross sections 
for the production of $n = 1, 2, \ldots$ pairs of quark-gluon strings, 
or cut Pomerons, and the cross section of double-diffractive process
\beq
\ds
\sigma_{ND}(s) = \sum \limits ^{\infty}_{n=1} \sigma_n (s) + 
\sigma_{DD} (s)\ .
\label{eq3}
\eeq
To find $\sigma_n (s)$, one can rely on the quasi-eikonal model
\cite{Kaid99,BTM76} which states that
\beqar
\ds
\label{eq4}
\sigma_{tot} (s) &=& \sum\limits _{n=0}^{\infty} \sigma_n (s) = 
\sigma_P\, f\left( \frac{z}{2} \right)\ , \\ 
\label{eq5}
\sigma_n (s) &=& \frac{\sigma_P}{n z} \left[ 1 - \exp{(-z)} 
\sum \limits ^{n-1}_{k=0} \frac{z^k}{k!} \right]\ ,\ k \geq 1 \\
\label{eq6}
\sigma_0 &=& \sigma_P\, \left[ f\left(\frac{z}{2} \right) - f(z)
\right]\ , \\
\label{eq7}
f(z) &=& \sum \limits ^{\infty}_{\nu = 1} \frac{(-z)^{\nu - 1}}{\nu
\nu !}\ . 
\eeqar
Here 
\beqar
\ds
\label{eq8}
\sigma_P &=& 8\pi \gamma_P \exp{( \Delta \xi )} \ , \\ 
\label{eq9}
z &=& \frac{2 C \gamma_P }{(R_P^2 + \alpha_P^\prime \xi)}\, 
\exp{( \Delta \xi )}\ .
\eeqar
The cross section $\sigma_0$ corresponds to diffraction contribution.
The parameters $\gamma_P$ and $R_P$ are Pomeron-nucleon vertex 
parameters, quantity $\Delta \equiv \alpha_P (0) - 1$, and 
$\alpha_P (0)$ and $\alpha_P^\prime$ are the intercept and the slope 
of the Pomeron trajectory, respectively. The quantity $C$ takes into 
account the deviation from the pure eikonal approximation $(C = 1)$ 
due to intermediate inelastic diffractive states, $\xi = \ln{(s/s_0)}$,
and $s_0$ is a scale parameter. 

For the diffractive processes displayed in Figs.~\ref{fig1}(b) and
\ref{fig1}(c), the fractions of momenta of initial hadrons carried 
by the sea quark pairs $x_{q \bar{q}}$ are determined according to 
distribution
\beq
\ds
u_{q \bar{q}}^h (x_{q \bar{q}}) \propto \frac{1 - x_{q \bar{q}}}
{x_{q \bar{q}}^{1 + \Delta}}
\label{eq10}
\eeq
Here we use a simple model in which the soft $q\bar{q}$-pair is produced
from a soft gluon emitted directly by valence quark (the so-called first
approximation). Thus, the proportionality coefficient in Eq.(\ref{eq10})
is not directly related to triple Pomeron vertex and should be fixed
from the comparison with experimental data.
The transverse momentum distribution of (anti)quarks in a proton in 
the low-mass excitation process shown in Fig.~\ref{fig1}(b) is given by
\beq
\ds
f_q (\vec{p_{\rm T}})\, d p_{\rm T} = \frac{b_1}{\sqrt{\pi}}\, 
\exp{(-b_1 p_{\rm T}^2)}\, d \vec{p_{\rm T}}\ ,
\label{eq11}
\eeq
where the slope parameter $b_1 = 20\, ({\rm GeV}/c)^{-2}$. Then, it is 
assumed that the valence (anti)diquark in the (anti)proton carries a
transverse momentum equal in magnitude and opposite in sign to the 
sum of transverse momenta of the other (anti)quarks. The number of
quark-gluon strings increases with collision energy; thus, the average 
transverse momentum of the (anti)diquark rises also.

The quantitative description of single-diffractive and 
double-diffractive processes at high energies was done in QGSM in terms 
of dressed triple-Reggeon and loop diagrams \cite{KP1,KP2}. The results 
obtained in \cite{KP1} for the cross sections of the diffractive 
processes are utilized in our MC model via the parametrizations
\beqar
\ds
\label{eq12}
\sigma_{SD}(s) &=& 0.68\, \left( 1 + \frac{36}{s} \right)\, 
\ln{ (0.6 + 0.2\,s) }\ , \\
\label{eq13}
\sigma_{DD}(s) &=& 1.65 + 0.27\,\ln{s}\ .
\eeqar  
Although these parameterisations are phenomenological, they agree
well with the asymptotics $\sigma_{D} \propto \ln{s}$ corresponding
to the Froissart bound, $\sigma_{tot} \propto (\ln{s})^2$.

\begin{figure}[htb]
 \resizebox{\linewidth}{!}{
\includegraphics{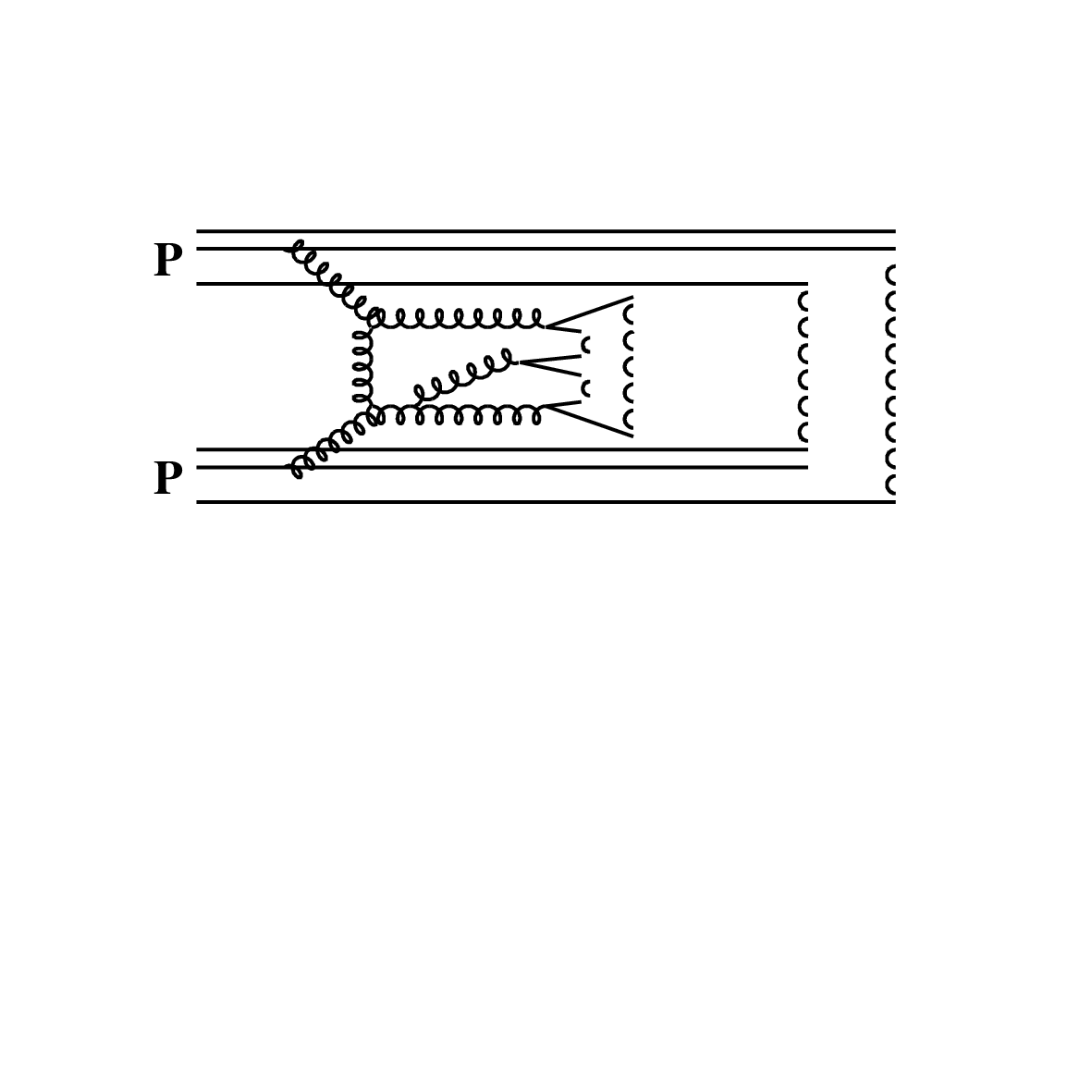}
}
\vspace{-4.0cm}
\caption{
String formation in hard gluon-gluon scattering and soft
Pomeron exchange in proton-proton collision.
}
\label{fig2}
\end{figure}

Soft processes dominate the particle production in hadronic 
interactions at intermediate energies. With the rise of the collision
energy, hard processes, resulting to formation of hadronic jets with
large transverse momenta, become important also. To take 
into account the jet formation and, on the other hand, to describe 
simultaneously the increase of the total and inelastic hadronic 
interaction cross section with rising $\sqrt{s}$, the eikonal model was 
properly modified in \cite{CTVK87} by introducing the new term that
represents the hard Pomeron exchange. The cut of the hard Pomeron 
leads to formation of two hadronic jets, see Fig.~\ref{fig2}. 
Therefore, the eikonal $u(s,b)$, that depends on the center-of-mass
energy $\sqrt{s}$ and the impact parameter $b$, can be decomposed onto 
the terms corresponding to soft and hard Pomeron exchange:
\beq
\ds
u(s,b) = u_{soft}(s,b) + u_{hard}(s,b)\ .
\label{eq14}
\eeq
The inelastic
hadronic cross section $\sigma_{inel}(s)$ is connected to the real 
part of the eikonal $u^R (s,b)$ as
\beq
\ds
\sigma_{inel}(s) = 2\pi \int \limits ^{\infty}_{0} \left\{ 1 -
\exp{\left[ - 2 u^R(s,b) \right]} \right\} \, b db \ .
\label{eq15}
\eeq
Recall that the concept of a (semi)hard Pomeron is nowadays a common
feature of all RFT-based MC models \cite{dpm,epos,phojet,ASC92,qgsjet2}
designed for the description of hadronic and nuclear interactions 
at ultrarelativistic energies. Other microscopic MC models also rely
on the picture of a soft+hard eikonal approach \cite{hijing}.

Following \cite{CTVK87,ASC92}, both soft and hard eikonals can be 
expressed as
\beq
\ds
u_{soft/hard}^R(s,b) = z_{soft/hard}(s)\, \exp{\left[- \frac{\beta^2}
{4\, \lambda_{soft/hard}(s)} \right]} \ ,
\label{eq16}
\eeq
where [cf. Eqs.~(\ref{eq4})$-$(\ref{eq9})]
\beqar
\ds
\label{eq17}
z_{soft/hard}(s) &=& \frac{\gamma_P}{\lambda_{soft/hard}(s)}\,
\left( \frac{s}{s_0} \right)^{\alpha_P (0) -1 } \\
\label{eq18}
\lambda_{soft/hard}(s) &=& R_P^2 + \alpha_P^\prime \ln{\left(
\frac{s}{s_0} \right)} \ .
\eeqar
Numerical values of the slopes and intercepts of the Pomeron 
trajectories and parameters of the hadron coupling to the Pomeron used 
in the model fit to experimental data are listed in Table~\ref{tab1}.
Note that these values deviate from the parameters of the soft and
hard Pomerons obtained in \cite{CTVK87,ICTV88} from the cross section of 
minijets measured by the UA1 Collaboration. To describe the LHC data at 
energies above $\sqrt{s} = 900$\,GeV, it was necessary to increase the 
soft Pomeron intercept to $\alpha_P(0) - 1 \approx 0.156$ and to 
increase the slope parameter $\alpha_P^\prime$ to 0.25.
\begin{table}
\caption{
\label{tab1}
Parameters of the soft and hard Pomerons used in the current version
of the QGSM.}
\vspace*{1ex}
\begin{ruledtabular}
\begin{tabular}{ccc}
Parameter         & Soft Pomeron & Hard Pomeron \\
\hline
$\alpha_P (0)$    &    1.15615    &  1.3217      \\ 
$\alpha_P^\prime$ &    0.25       &  0           \\ 
$\gamma_P$        &    1.27475    &  0.021       \\ 
$R_P$             &    2.0        &  2.4         \\ 
\end{tabular}
\end{ruledtabular}
\end{table}

For the hard Pomeron, the fractions of momenta of the gluons are
generated from the structure function \cite{Baier_struc}:
\beq
\ds
\label{eq:gluon_dist}
xG(x,Q^2)=C_g(\bar{s})\, x^{\eta_1^g(\bar{s})}(1-x)^{\eta_2^g(\bar{s})}\,,
\eeq
with 
\beq
\ds
\bar{s}=\ln{[(\ln{Q^2}/\Lambda)/(\ln{Q_0^2}/\Lambda)]}\,,\quad 
\Lambda = 200~{\rm MeV}
\label{eq20}
\eeq
and
\beqar
\ds
C_g(\bar{s})&=&2.01 - 3.56\bar{s}+1.98\bar{s}^2\\
\label{eq21}
\eta_1^g(\bar{s})&=&-1.13\bar{s}+0.48\bar{s}^2\\
\label{eq22}
\eta_2^g(\bar{s})&=&2.9+0.813\bar{s} \ .
\label{eq23}
\eeqar
The transverse momentum is generated from the distribution
\beq
\ds
\label{pt_hard}
f(p_{\rm T})dp_{\rm T} = \alpha(1 + p_{\rm T})^\beta \ ,
\eeq
where $\alpha$ and $\beta$ are determined for each event by fitting the
summed cross sections (calculated from Ref.~\cite{FFF_crosec} for 
$y_1=y_2=0$) for all $gg\to gg$ and $gg \to q\bar{q}$ processes, i.e., 
for all hard Pomerons, to an envelope  
\beq
\ds
\label{crosec_hard}
\frac{d^3 \sigma}{dp_{\rm T}^2 dy_1 dy_2} \lesssim 
\alpha(1 + p_{\rm T})^\beta.
\eeq
The $p_{\rm T}$ values are then generated within the following limits:
\beqar
\ds
p_{{\rm T},min}(s)&=&p_{{\rm }T,0}+0.0054\, s^{0.31393} \\
\label{eq26}
p_{{\rm T},max}(s)&=&p_{{\rm T},min}(s)+6.0+0.08\, s^{\alpha_P^{hard}(0)-1}
\label{eq27}
\eeqar
This procedure generates an explicit dependence of the transverse 
momentum of the produced particles on the collision energy $\sqrt{s}$. 
As $\sqrt{s}$ increases, more and more hard Pomerons emerge. The 
differential cross section given by Eq.(\ref{crosec_hard}) increases, 
rendering the power-law distribution for $p_{\rm T}$ harder, with 
additionally increased lower and upper cut-off values for the 
distribution.

Then, the AGK cutting rules enable one to express the inelastic 
cross section as
\beq
\ds
\nonumber 
\sigma_{inel} (s) = \sum \limits_{i,j = 0; i+j \geq 1}^{ }
\sigma_{ij}(s)\ ,
\label{eq28}
\eeq
where 
\beqar
\ds
\sigma_{ij}(s) &=& 2 \pi \int \limits_{0}^{\infty} b db\,
\exp{\left[ -2 u^R(s,b) \right]}\\
\nonumber
 &\times & \frac{\left[ 2u^R_{soft}(s,b)  \right] ^i}{i !}
\frac{\left[ 2u^R_{hard}(s,b)  \right] ^j}{j !} \ .
\label{eq29}
\eeqar
The last expression can be used to determine the number of quark-gluon
strings and hard jets via the number of cut soft and hard Pomerons,
respectively. At very high energies, one has to take into account the
effects of shadowing of partonic distributions both in nucleons and
in nuclei. In the Reggeon calculus such processes correspond to the 
so-called enhanced diagrams \cite{enh_diag} describing the 
interactions between Pomerons. These diagrams are not implemented yet 
in the current MC version of the QGSM.  

As has been discussed in the literature (see e.g. 
Refs.~\cite{AGK-viol1,AGK-viol2,AGK-viol3}), the AGK cutting rules are 
violated for multiple gluon production from Pomeron vertices. In our 
model, however, the number of gluons produced from single hard Pomeron 
vertex is limited to two. Therefore, an exchange of a hard Pomeron 
leads only to $gg \to gg$ or $gg \to q\bar{q}$ processes, i.e. only a 
double gluon emission from a hard Pomeron may happen. As was pointed 
out in \cite{AGK-viol2}, the AGK cutting rules provide the leading 
contribution for the inclusive double-gluon emission process.  

The multi-Pomeron exchanges become very
important with increasing c.m. energy of hadronic collision. For 
instance, the contribution of a single-cylinder diagram to the 
scattering amplitude is proportional to $(s/s_0)^{\alpha_P(0) - 1},\ 
\alpha_P(0) > 0$. In contrast, the contributions coming from the 
$n$-Pomeron exchanges grow as $(s/s_0)^{n \Delta}$. Although in the 
framework of the $1/N$-expansion the $n$-Pomeron exchange amplitudes 
are suppressed by factor $1/N^{2n}$, the quickly rising term 
$s^{n \Delta}$ dominates over the suppression factor at 
ultrarelativistic energies.

There is no unique theoretical prescription for modeling the
fragmentation of a string with a given mass, momentum and quark 
content into hadrons. In the presented version of the QGSM the
Field-Feynman algorithm \cite{FF_frag} is employed. It enables one to
consider emission of hadrons from both ends of the string with equal
probabilities. The breakup procedure invokes the energy-momentum
conservation and the preservation of the quark numbers. 
The transverse momentum of the (di)quarks in the vacuum pair is
determined by the power-law probability
\beqar
\ds
f(p_{\rm T}^2)\, d p_{\rm T}^2 &=& \frac{3\,D\, b_2(s)}{\pi \, 
(1 + D p_{\rm T}^2)^4}\, dp_{\rm T}^2\ ,\\
\label{eq30}
b_2(s) &=& 0.325 + 0.016\, \ln{s}\ ,
\label{eq31}
\eeqar
with $D = 0.34$\, (GeV/$c$)$^{-2}$.
 
Further details of the MC version of QGSM and its extension to $h+A$ 
and $A+A$ collisions can be found in \cite{mc_qgsm,ber_tue,ell_fl}.

\section{Comparison with data and predictions for LHC}
\label{sec3}

\subsection{Cross sections}
\label{sigmas}

For the comparison with model results concerning the pseudorapidity
and transverse momentum distributions we used experimental data 
obtained by the UA5 Collaboration for antiproton-proton collisions
at c.m. energies $\sqrt{s} = 200$\,GeV, 546\,GeV and 900\,GeV
\cite{UA5_rep}, by the UA1 Collaboration for $\bar{p}p$ collisions
at $\sqrt{s} = 546$\,GeV \cite{UA1}, by the CDF and the E735 
Collaborations for $\bar{p} p$ collisions at $\sqrt{s} = 1800$\,GeV 
\cite{CDF,E735}, and recent CERN LHC data obtained for $pp$ 
interactions at $\sqrt{s} = 900$\,GeV, 2360\,GeV, 7\,TeV, 8\,TeV and 
13\,TeV by the ALICE Collaboration
\cite{ALICE_1,ALICE_2,ALICE_3,ALICE:2013bva,Adam:2015pza},
by the CMS Collaboration
\cite{CMS_1,CMS_2,Khachatryan:2015jna}, 
and by the TOTEM Collaboration \cite{Chatrchyan:2014qka}. 
At such high energies, the annihilation cross section is almost zero and 
the main characteristics of particle production in $pp$ and $\bar{p} p$ 
interactions are essentially similar.

\begin{figure}[htb]
 \resizebox{\linewidth}{!}{
\includegraphics{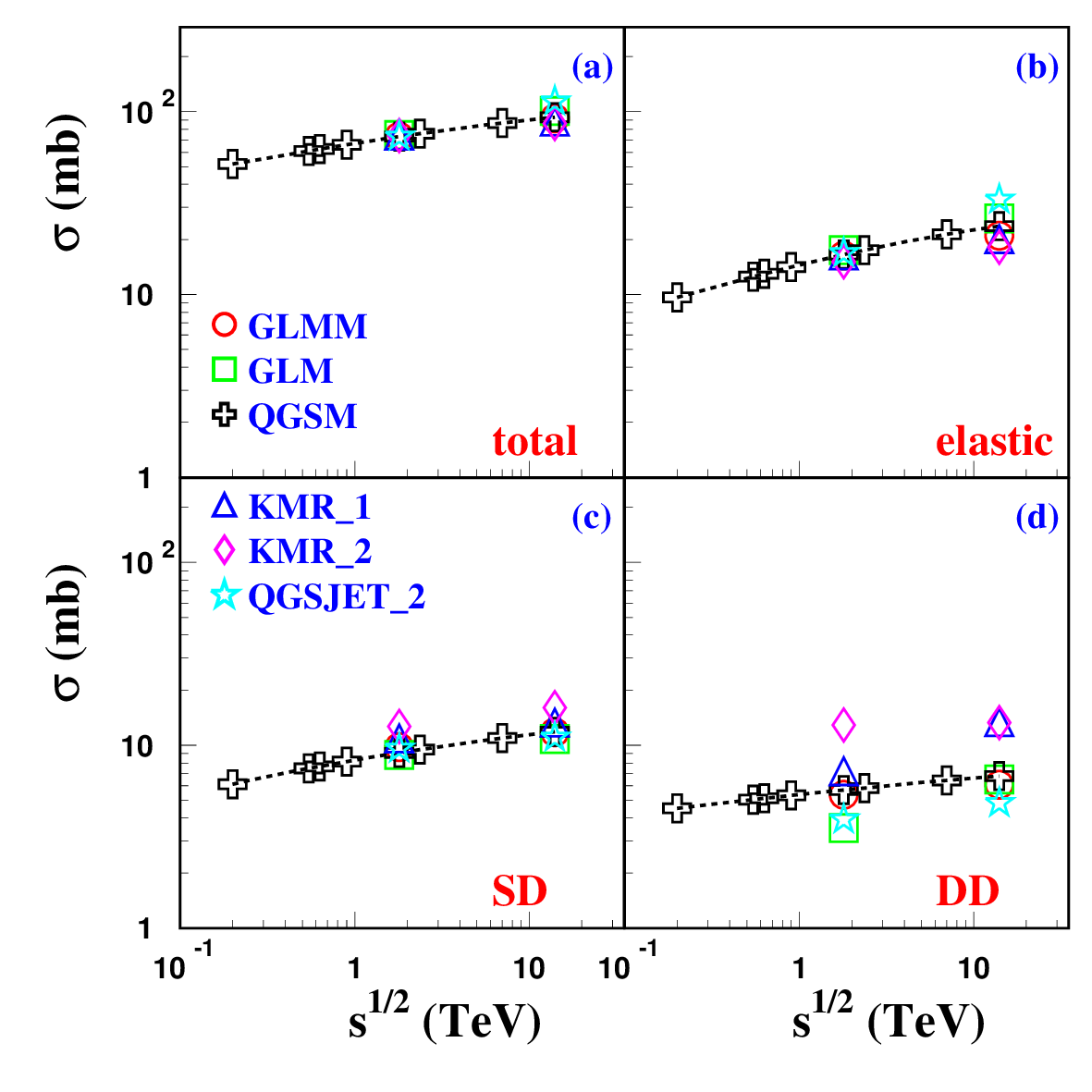}
}
\caption{
(Color online)
(a) Total, (b) elastic, (c) single-diffractive and (d) 
double-diffractive cross sections as functions
of $\sqrt{s}$ obtained in the models GLMM (circles) \cite{GLMM}, GLM
(squares) \cite{GLM}, QGSM (crosses), KMR-1 (triangles) \cite{KMR_1},
KMR-2 (diamonds) \cite{KMR_2}, and QGSJET-2 (stars) \cite{qgsjet2}. 
Dashed line, connecting the QGSM points, is drawn to guide the eye. }
\label{fig3}
\end{figure}

Total and elastic cross sections are listed in Table~\ref{tab2} 
together with the cross sections of multichain, single-
and double-diffraction processes for energies ranging from $\sqrt{s} 
= 200$\,GeV to $\sqrt{s} = 14$\,TeV.
\begin{table}
\caption{
\label{tab2}
Total, elastic, multichain, single-diffraction and double-diffraction
cross sections of $pp$ collisions calculated by the QGSM.}
\vspace*{1ex}
\begin{ruledtabular}
\begin{tabular}{cccccc}
$\sqrt{s}$ (GeV) & $\sigma_{tot}$ (mb) & $\sigma_{el}$ (mb) & 
$\sigma_{P}$ (mb) & $\sigma_{SD}$ (mb) & $\sigma_{DD}$ (mb) \\
\hline 
  200  &  51.62  &   9.67  &  31.12  &   6.12  &   4.51   \\ 
  546  &  60.83  &  12.51  &  35.72  &   7.48  &   5.05   \\ 
  630  &  62.25  &  12.97  &  36.42  &   7.67  &   5.13   \\
  900  &  65.85  &  14.15  &  38.19  &   8.16  &   5.32   \\ 
 1800  &  72.97  &  16.55  &  41.61  &   9.10  &   5.70   \\  
 2360  &  75.74  &  17.50  &  42.92  &   9.47  &   5.84   \\  
 7000  &  86.60  &  21.31  &  47.91  &  10.95  &   6.43   \\  
14000  &  93.07  &  23.61  &  50.76  &  11.89  &   6.80  
\end{tabular}
\end{ruledtabular}
\end{table}
Compared to those at $\sqrt{s} = 900$\,GeV, $\sigma_{tot}$, 
$\sigma_{el}$ and $\sigma_{SD}$ increase at $\sqrt{s} = 14$\,TeV by 
nearly 50\%, whereas $\sigma_{DD}$ increases by less than 30\%.
For better understanding of theoretical uncertainties,
the results obtained for the $\sigma_{tot},\ \sigma_{el},\ 
\sigma_{SD}\ {\rm and}\ \sigma_{DD}$ are compared in Fig.~\ref{fig3}
with the available predictions of other models 
\cite{qgsjet2,GLMM,GLM,KMR_1,KMR_2}, which also rely on the RFT. We 
see that for Tevatron energy $\sqrt{s} = 1.8$\,TeV all models agree 
within 5\% accuracy limit for all but double-diffraction cross 
section. At top LHC energy $\sqrt{s} = 14$\,TeV the predictions for 
$\sigma_{tot}$ and $\sigma_{SD}$ are still close to each other,
whereas the Durham models, KMR-1 and KMR-2, predict 50\% excess of
$\sigma_{DD}$ compared to other models. Results of the present version 
of QGSM are close to the calculations of the GLMM model \cite{GLMM}. 
On the other hand, QGSM is alike to QGSJET model \cite{qgsjet2}, which 
also contains soft and hard Pomerons with the parameters similar to 
those listed in Table~\ref{tab1} except of the Pomeron slopes. QGSJET 
yields larger total and elastic cross sections at 14~TeV. A discussion 
of the similarities and differences between the models presented here 
can be found in \cite{GLM}. 

Inelastic and diffractive cross sections have been measured at the 
LHC in \cite{Abelev:2012sea,Antchev:2013gaa,Aaij:2014vfa}. The results 
are listed in Table~\ref{tab3}. After comparison of experimental data
with the QGSM calculations from Table~\ref{tab2}, it turns out that the 
model works reasonably well. It tends to slightly underestimate most 
of the cross sections, although, e.g., the inelastic cross section in 
the model is quite close to the one reported by the LHCb Collaboration
\cite{Aaij:2014vfa}.
\begin{table}
\caption{
\label{tab3}
Inelastic, elastic, single-diffractive and double-diffractive
cross sections of $pp$ collisions measured at the LHC.}
\vspace*{1ex}
\begin{ruledtabular}
\begin{tabular}{ccccc}
$\sqrt{s}$ (TeV)  & $\sigma_{inel}$ (mb) & 
$\sigma_{el}$ (mb)&$\sigma_{SD}$ (mb) & $\sigma_{DD}$ (mb) \\
\hline 
0.9 (ALICE) \cite{Abelev:2012sea}  &52.5 &   &11.2&5.6       \\
2.76 (ALICE)                       &62.8 &   &12.2&7.8       \\
7.0 (ALICE)                        &73.2 &   &14.9&9.0       \\
7.0 (LHCb) \cite{Aaij:2014vfa}     &66.9 &   &    &          \\ 
7.0 (TOTEM) \cite{Antchev:2013gaa} &73.15& 25.43& &       
\end{tabular}
\end{ruledtabular}
\end{table}

\subsection{Transverse momentum spectra}
\label{p_t}

\begin{figure}[htb]
 \resizebox{\linewidth}{!}{
\includegraphics{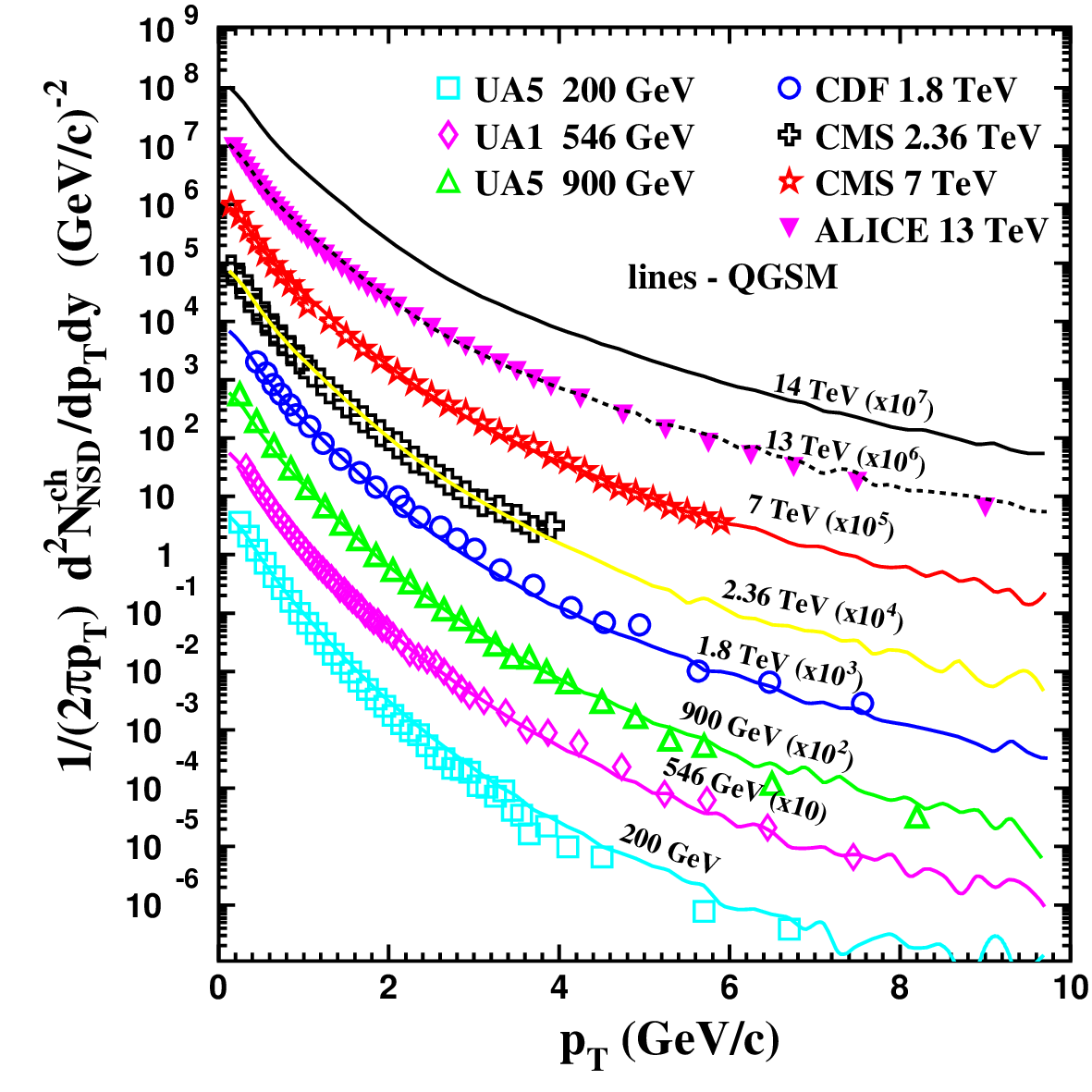}
}
\caption{
(Color online)
Transverse momentum distributions of the invariant cross section of
charged particles in NSD $pp$ collisions obtained in QGSM at 
$|y| \leq 2.5$ for all energies in question. Experimental data are 
taken from \cite{UA5_rep,UA1,CDF,CMS_2,Adam:2015pza}.
}
\label{fig4}
\end{figure}
\begin{figure}[htb]
 \resizebox{\linewidth}{!}{
\includegraphics{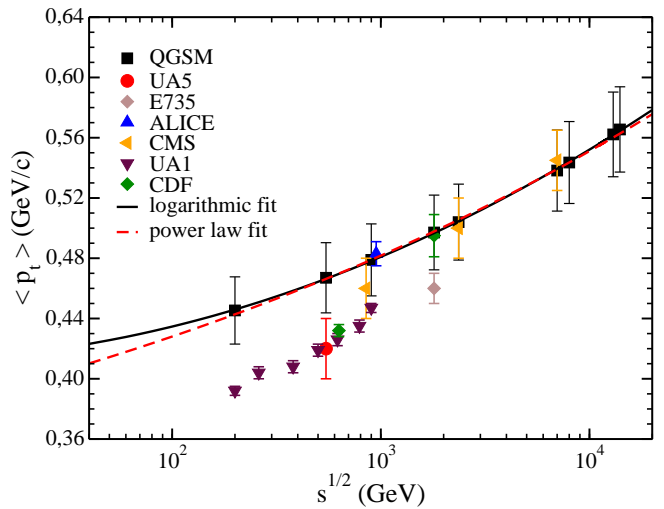}
}
\caption{
(Color online)
Average transverse momentum as a function of $\sqrt{s}$. Squares 
present the QGSM results; other symbols denote experimental data
from \protect\cite{UA5_rep,UA1,CDF,E735,ALICE_3,CMS_2}. 
Solid and dashed lines are fit to logarithmic and power-law 
dependences, respectively. See text for details.
}
\label{fig5}
\end{figure}

The transverse momentum distributions of the invariant cross section
$\ds {\rm E} \frac{d^3 \sigma}{ d p^3}$ divided to $\sigma_{tot}$ for 
charged particles in nonsingle-diffractive (NSD) $pp$ collisions
at all energies in question are presented in Fig.~\ref{fig4}. We see 
that the QGSM reproduces the experimental data in a broad energy range
pretty well. The spectra become harder with increasing $\sqrt{s}$; 
thus, the average transverse momentum of produced hadrons should grow
also. Figure~\ref{fig5} displays the $\langle p_{\rm T} \rangle$ of 
charged particles in NSD $pp$ events calculated in QGSM and compared 
to experimental data. We assume here 5\% systematic errors for the 
extraction of mean $p_{\rm T}$ because we do not apply any extrapolation 
procedure to the generated spectra, as it is usually done in the 
experiments. Results of the fit of model simulations to quadratic 
logarithmic dependence and to power-law dependence are as follows: 
\beqar
\ds
\nonumber
\langle p_{\rm T} \rangle &=& 0.417 -0.0035 \ln{s} + 0.00059 \ln^2{s}\ , 
\\ \nonumber
\langle p_{\rm T} \rangle &=& 0.243 + 0.12 {\rm E}^{0.1107}\ .
\eeqar
In the last expression ${\rm E} = \sqrt{s}/2$, and the exponent 0.1107 
is not a free parameter. According to \cite{MLP10}, this exponent is 
just half of the exponent of the power-law fit to $d N/d\eta$ 
distribution (see below). As one can see in Fig.~\ref{fig5}, the 
difference between the two parametrizations of mean $p_{\rm T}$ is 
negligible even for top LHC energy $\sqrt{s} = 14$\,TeV. 

\begin{figure}[htb]
 \resizebox{\linewidth}{!}{
\includegraphics{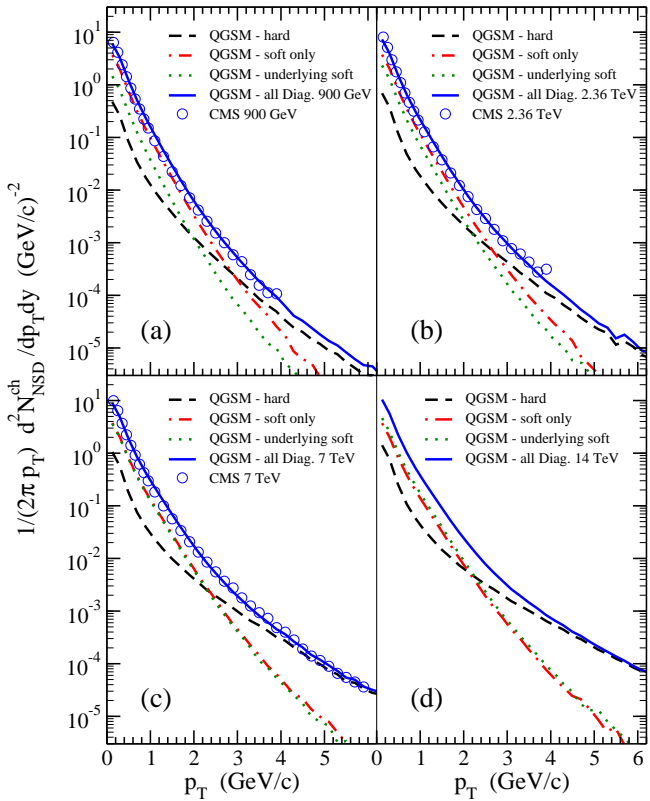}
}
\caption{
(Color online)
Transverse momentum distributions of the invariant cross section of
charged particles in NSD $pp$ collisions 
at (a) $\sqrt{s} = 900$\,GeV, (b) $\sqrt{s} = 2360 $\,GeV, (c) 
$\sqrt{s} = 7 $\,TeV and (d) $\sqrt{s} = 14$ \,TeV calculated in QGSM. 
Combined contribution of all processes and, separately, of only soft, 
hard and underlying soft subprocesses are shown by solid, dash-dotted,
dashed and dotted lines, respectively (see text for details). 
Experimental data plotted in panels (a), (b) and (c) are taken from 
\cite{CMS_1,CMS_2}.
}
\label{fig6}
\end{figure}

To study the interplay between the soft and hard processes, 
we show separately in Fig.~\ref{fig6} their fractional contributions 
and combined results for $pp$ collisions at $\sqrt{s} = 900$\,GeV, 
2.36\,TeV, 7\,TeV and at top LHC energy $\sqrt{s} = 14$\,TeV. Moreover, 
the $p_{\rm T}$ dependence of the underlying soft processes from 
the collisions with at least one hard Pomeron is displayed in these 
plots as well as with the experimental data of the CMS 
Collaboration. As expected, the soft processes dominate at low and 
intermediate transverse momenta, whereas at higher transverse momenta 
the major contribution to the cross section comes from the minijets. 
The crossover between the hard and soft branches takes place at 
$p_{\rm T} \approx 2.8$\,GeV/$c$ for the reactions at $\sqrt{s} = 
900$\,GeV. It is shifted to $p_{\rm T} \approx 2.2$\,GeV/$c$ at 
$\sqrt{s} = 14$\,TeV. The slopes of the $p_{\rm T}$ spectra for both 
soft and underlying soft processes are similar. At $\sqrt{s} = 7$\,TeV 
and 14\,TeV, both lines coincide; i.e., the contributions to the 
invariant cross sections from barely soft Pomeron processes are equal 
to those from the soft Pomerons exchanges, accompanied by one or more 
hard Pomeron ones.

\vspace{0.5cm}

In view of these results it becomes clear, what process generates 
the growing mean $p_{\rm T}$ in our model. Particle production from 
soft and hard Pomerons includes different distributions for the 
transverse momentum. Their relative contributions are energy dependent, 
see Fig. \ref{fig6}. 
Additionally, both distributions depend explicitly on the collision 
energy. With growing $\sqrt{s}$, more and more hard Pomerons
are exchanged, rendering the spectra of secondaries harder.

\subsection{Rapidity distributions}
\label{rapidity}

Let us briefly recall the main assumptions and predictions of the
hypothesis of Feynman scaling \cite{Feyn_scal}. It requires scaling
behavior of particle spectra within the whole kinematically allowed
region of the Feynman scaling variable $x_{\rm F} \equiv p_{||} / 
p_{||}^{max}$ or, alternatively, c.m. rapidity $y^\ast$ at 
ultrarelativistic energies $s \rightarrow \infty $. In addition, the
existence of nonvanishing central area $|x_{\rm F} | \leq x_0\, ,\ 
x_0 \sim 0.1 $ is postulated. In terms of rapidity this central region 
increases with rising $\sqrt{s}$ as
\beq
\ds
(\Delta y^\ast)_{centr} \approx 2\, \ln{ \left[ x_0 \sqrt{s} / m_{\rm T} 
\right] }
\label{eq32}
\eeq
provided the transverse mass $m_{\rm T} = \sqrt{m_0^2 + p_{\rm T}^2}$ 
is finite. In contrast, the fragmentation region remains constant
\beq
\ds
(\Delta y^\ast)_{frag} \approx \ln{ (1/x_0) }\ .
\label{eq33}
\eeq
From here, it follows that (i) in the central area the particle density
$\rho_{cent} (y^\ast, p_{\rm T}, s)$ depends on neither $y^\ast$ nor 
$\sqrt{s}$, i.e., $\rho_{cent} \equiv \rho_{cent} (p_{\rm T})$, and 
rapidity spectra of particles have, therefore, a broad plateau; (ii) 
this area gives a main contribution to average multiplicity of produced 
hadrons; (iii) contribution to the average multiplicity from the 
fragmentation regions is energy independent.

\begin{figure}[htb]
 \resizebox{\linewidth}{!}{
\includegraphics{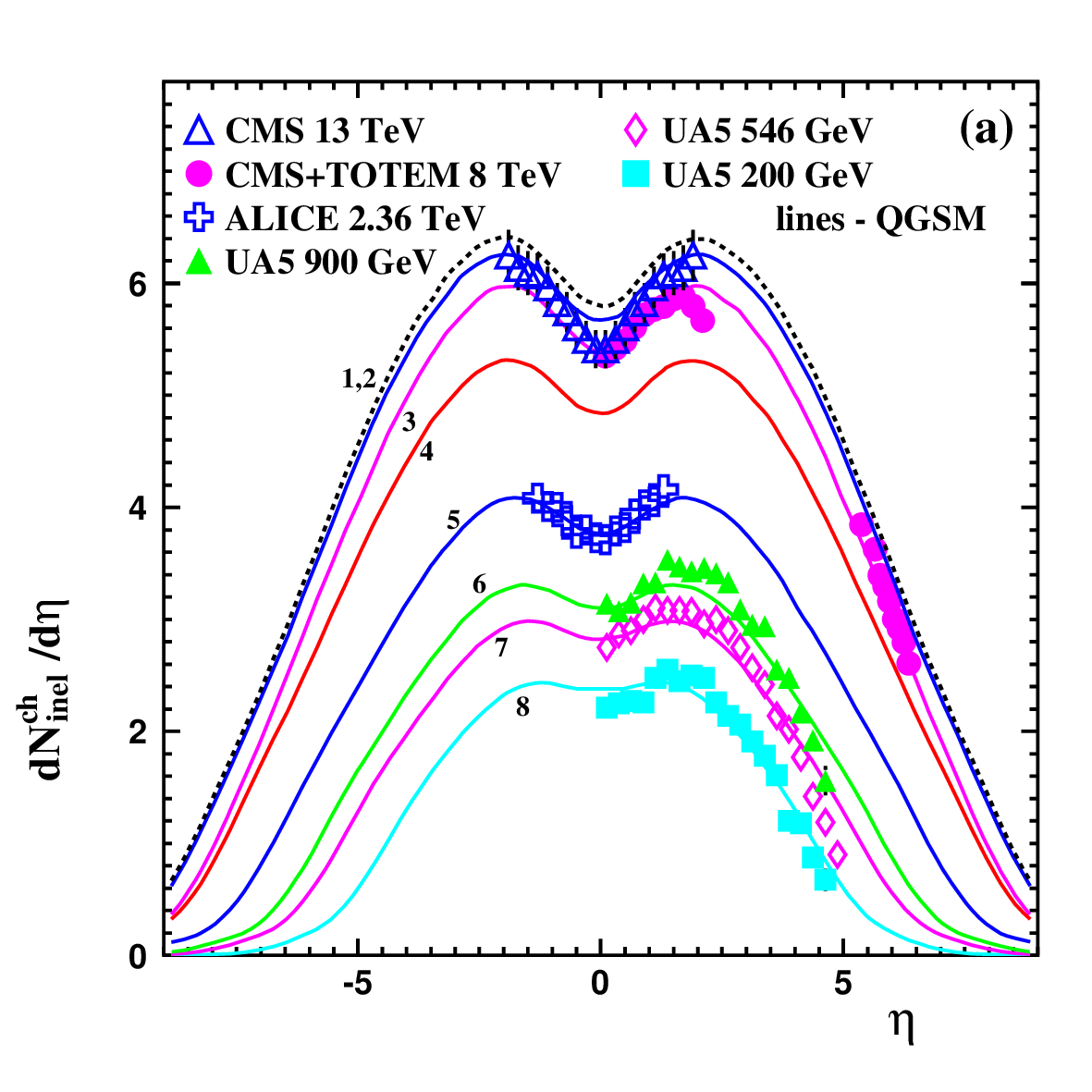}}
 \resizebox{\linewidth}{!}{
\includegraphics{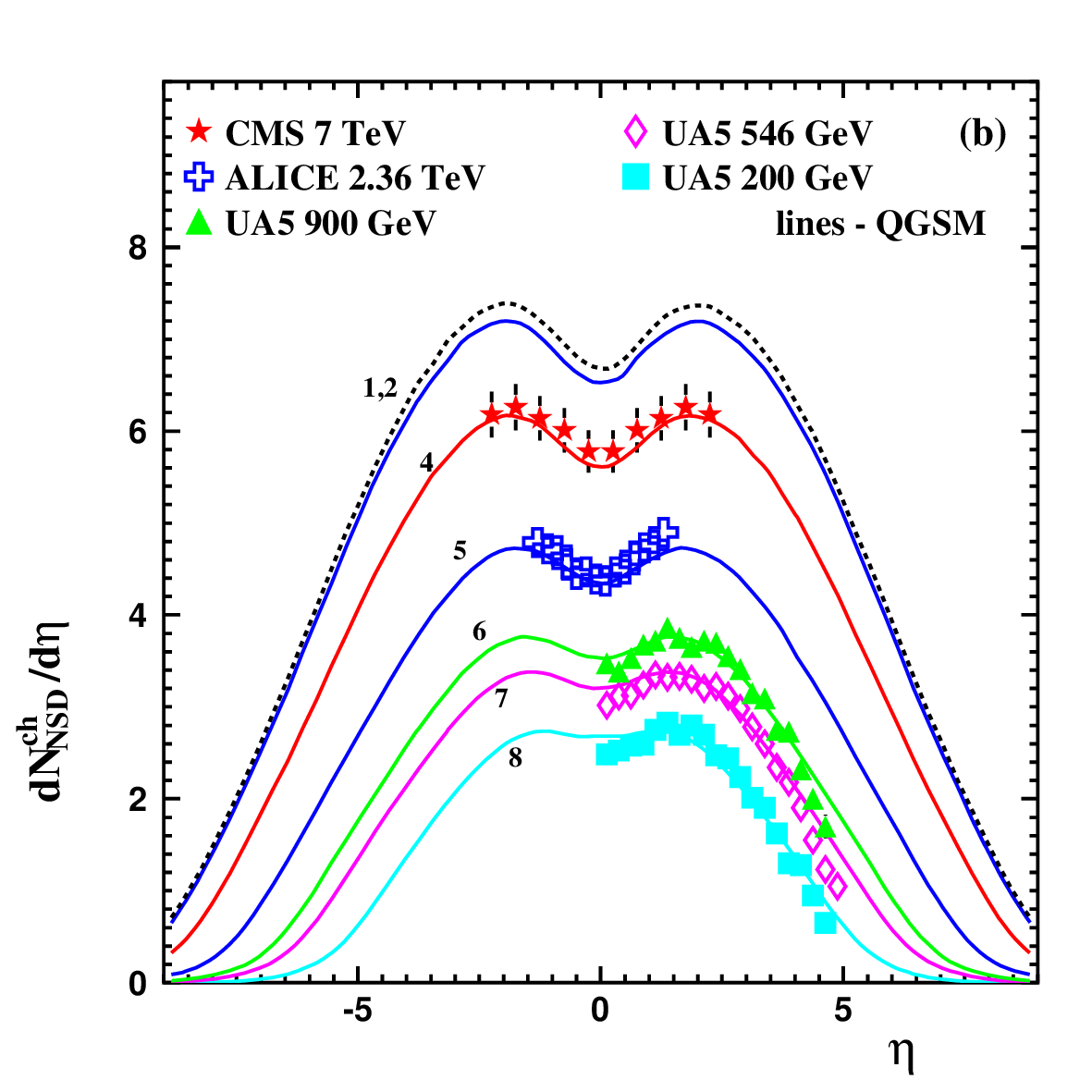}
}
\caption{
(Color online)
The charged particle pseudorapidity spectra for (a) inelastic and 
(b) nonsingle-diffractive events calculated in QGSM for $pp$ 
collisions at $\sqrt{s}= 200$\,GeV (8), 546\,GeV (7), 
900\,GeV (6), 2.36\,TeV (5), 7\,TeV (4), 8\,TeV (3), 13\,TeV (2), and 
14\,TeV (1). Data are taken from 
\cite{UA5_rep,ALICE_2,CMS_2,Khachatryan:2015jna,Adam:2015pza}.
}
\label{fig7}
\end{figure}
The charged particle pseudorapidity spectra 
$\ds \frac{1}{\sigma_{inel}} \frac{d \sigma_{inel}}{d \eta}$ and 
$\ds \frac{1}{\sigma_{NSD}}\frac{ d \sigma_{NSD}}{d \eta}$ for 
inelastic and nonsingle-diffractive
events, respectively, are displayed in Figs.~\ref{fig7}(a) and
\ref{fig7}(b) together with the $pp (\bar{p} p)$ data at 
$\sqrt{s} = 200$\,GeV, 546\,GeV, 900\,GeV, 2.36\,TeV, 
7\,TeV and 13\,TeV. 
QGSM predictions for $\sqrt{s} = 14$\,TeV are plotted here also. The 
model gives a good description of these distributions within the 
indicated energy range except, maybe, a not very distinct dip at 
midrapidity for the lowest energy in question $\sqrt{s} = 200$\,GeV. 
For $pp$ collisions at top LHC energy, QGSM predicts a further increase 
of the central particle densities to
$$ 
\ds
\left. \frac{d N_{inel}}{d \eta} \right\vert_{\eta = 0} = 5.8 
\quad , \quad
\left. \frac{d N_{NSD}}{d \eta} \right\vert_{\eta = 0} = 6.7 \ .
$$ 
Compared to the $\sqrt{s} = 7$\,TeV, the rise of the central particle 
density at $\sqrt{s} = 14$\,TeV is expected to be about 20\%.

\begin{figure}[htb]
 \resizebox{\linewidth}{!}{
\includegraphics{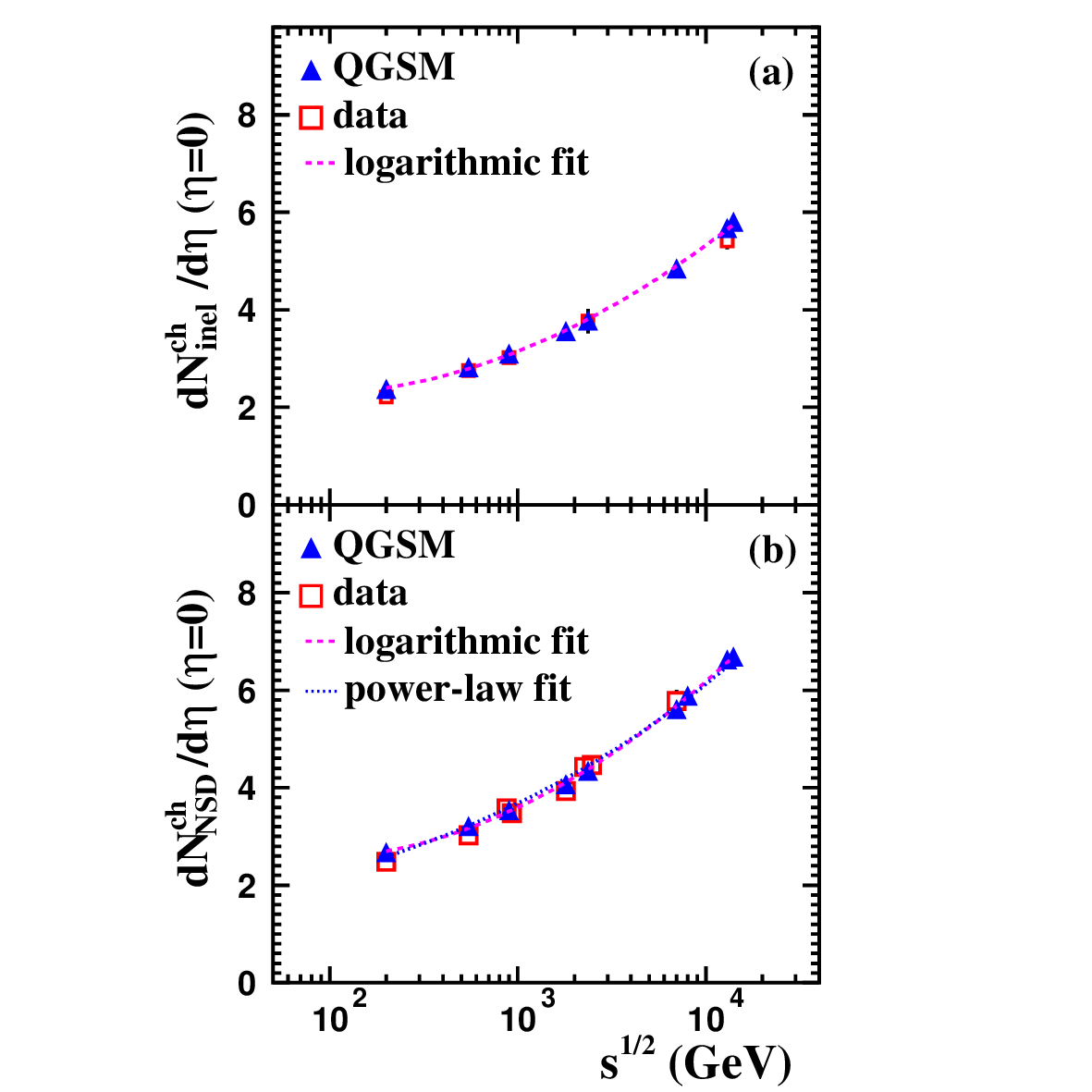}
}
\caption{
(Color online)
The charged particle density at midrapidity as a function of 
$\sqrt{s}$ for (a) inelastic and (b) nonsingle-diffractive 
collisions. Dashed lines show the results of the fit to expression
$a + b\,\ln{s} + c\,\ln^2{s}$, dotted line (b) $-$ to the power-law
dependence $d\, \sqrt{s}^\lambda$.
}
\label{fig8}
\end{figure}
In Fig.~\ref{fig8}, the charged particle density at $\eta = 0$ is
presented as a function of the c.m. energy $\sqrt{s}$ for inelastic
(upper plot) and nonsingle-diffractive (bottom plot) events. The
experimental data for inelastic collisions below $\sqrt{s} = 546$\,GeV
are well described by a linear dependence on $\ln{s}$ \cite{UA5_rep}.
The striking evidence of the first LHC results for $pp$ interactions at
$\sqrt{s} = 900$\,GeV, 2.36\,GeV and 7\,TeV is the quadratic dependence 
of the increase of midrapidity density of charged particles with rising 
$\ln{s}$ \cite{CMS_2}. The theory of CGC suggests a power-law rise 
\cite{Lev10,MLP10}. In the QGSM these trends hold also, and the 
fitting parametrizations for c.m. energies from 200\,GeV to 14\,TeV are
$$
\ds
\left. \frac{d N_{inel}}{d \eta} \right\vert_{\eta = 0} (s) = 
4.36 - 0.507\,\ln{s} + 0.03\,\ln^2{s} \ ,
$$
$$
\ds
\left. \frac{d N_{NSD}}{d \eta} \right\vert_{\eta = 0} (s) = 
5.015 - 0.60\,\ln{s} + 0.036\,\ln^2{s} \ ,  
$$
$$
\ds
\left. \frac{d N_{NSD}}{d \eta} \right\vert_{\eta = 0} (s) = 
0.77\,E^{0.22} \ .
$$
As in the mean $p_{\rm T}$ case, there is a hair's width difference 
between the two curves representing the logarithmic and the power-law 
fit, respectively. Indicating a further increase of particle density at 
$\eta = 0$ with rising energy, the model favors violation of the 
Feynman scaling at midrapidity; otherwise, the particle density there 
should not depend on $\sqrt{s}$.

It is interesting to compare the QGSM predictions for the charged
particle multiplicity in $pp$ collisions at LHC with that obtained
by the extrapolation of pseudorapidity distributions measured at
lower energies. This method \cite{Busza08} employs the energy
independence of the slopes of the pseudorapidity spectra combined with
logarithmic proportionality to $\sqrt{s}$ of both the width and the 
height of the distributions. Therefore, any data set from 
Figs.~\ref{fig7}(a) and \ref{fig7}(b) can be used for the 
extrapolation, and the results are \cite{Busza08}
$$ 
\ds
\left. \frac{d N_{NSD}}{d \eta} \right\vert_{\eta = 0} = 4.6 \pm 0.4 \ ,
\quad 
\left. \frac{d N_{NSD}}{d \eta} \right\vert_{\eta = \pm 2} = 
5.25 \pm 0.7 \ . 
$$ 
These predictions are significantly lower than the recent 
experimental data from LHC and the QGSM calculations. 

\begin{figure}[htb]
 \resizebox{\linewidth}{!}{
\includegraphics{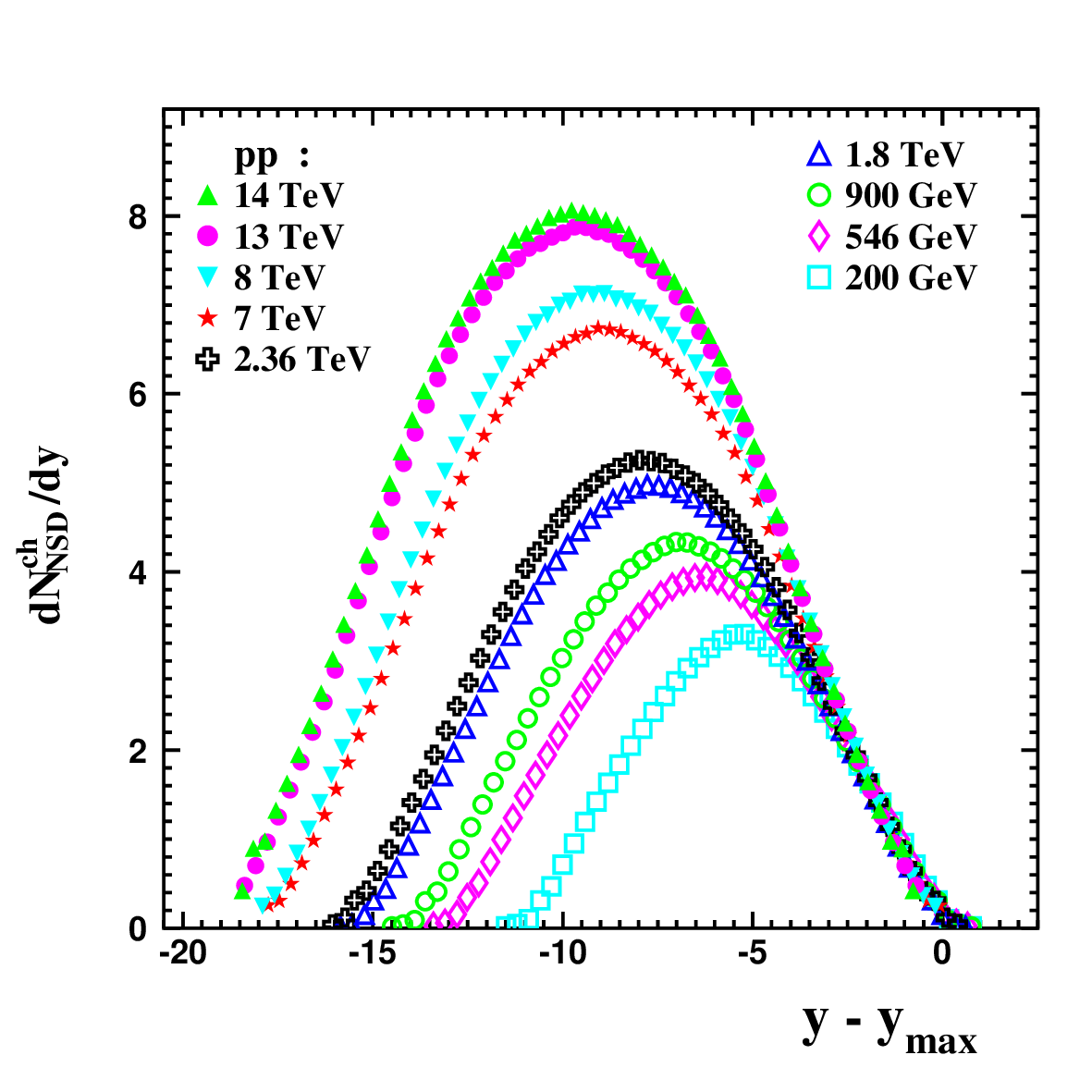}
}
\caption{
(Color online)
The distributions 
$\ds \frac{1}{\sigma_{NSD}}\, \frac{d \sigma_{NSD}}{d y}$ as functions 
of rapidity difference $y - y_{max}$ obtained in QGSM for energies
$\sqrt{s} = 200$\,GeV, 546\,GeV, 900\,GeV, 1.8\,TeV, 2.36\,TeV, 
7\,TeV, 13\,TeV and 14\,TeV.
}
\label{fig9}
\end{figure}
Another feature that is closely related to Feynman scaling is the 
so-called extended longitudinal scaling \cite{ext_long_scal} 
exhibited by the slopes of (pseudo)rapidity spectra. In the QGSM
these slopes are identical in the fragmentation region $y_{beam}
\geq -2.5$ as shown in Fig.~\ref{fig9}, where the distributions
$\ds \frac{1}{\sigma_{NSD}}\, \frac{d \sigma_{NSD}}{d y}$ are
expressed as functions $y - y_{max}$. QGSM indicates that the
extended longitudinal scaling remains certainly valid at LHC. This 
result contradicts to the prediction based on the statistical 
thermal model \cite{CST_08}. The latter fits the measured rapidity 
distributions to the Gaussian, extracts the widths of the Gaussians 
and implements the energy dependence of the obtained widths to 
simulate the rapidity spectra at LHC. The extrapolated distribution 
was found to be much narrower \cite{CST_08} compared to that presented 
in Fig.~\ref{fig9}. We are eagerly awaiting the LHC measurements of 
$pp$ collisions in the fragmentation regions to resolve the obvious 
discrepancy. Note that experimentally the extended longitudinal 
scaling was found to hold to 10\% in a broad energy range from 
$\sqrt{s} = 30.8$\,GeV to 900\,GeV \cite{UA5_rep}.

\begin{figure}[htb]
 \resizebox{\linewidth}{!}{
\includegraphics{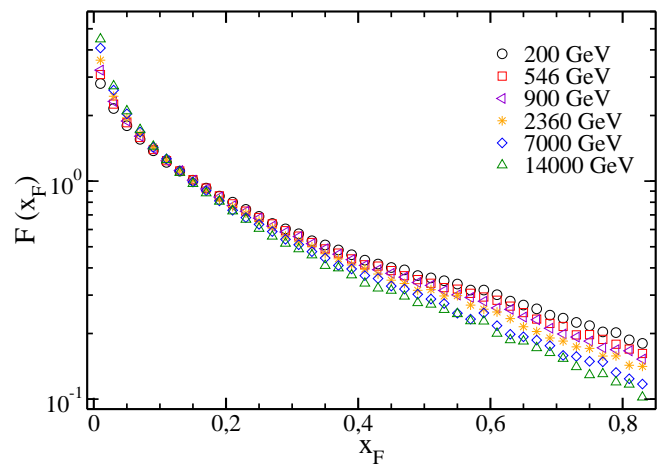}
}
\caption{
(Color online)
The invariant distribution $F(x_{\rm F})$ in nonsingle-diffractive $pp$
collisions obtained in QGSM at $\sqrt{s} = 200$\,GeV, 546\,GeV, 
900\,GeV, 1.8\,TeV, 2.36\,TeV and 14\,TeV.
}
\label{fig10}
\end{figure}

The emergence of the extended longitudinal scaling as well as Feynman
scaling in the QGSM is not accidental. It arises due to short range 
correlations in rapidity space. The correlation function of particle
$i$ and particle $j$, produced in the string fragmentation, drops 
exponentially with rising rapidity difference
\beqar
\ds
\nonumber
C(y_i,y_j) &=& \frac{d^2 \sigma}{\sigma_{inel}\, d y_i d y_j} -
\frac{d \sigma}{\sigma_{inel}\, d y_i} \, 
\frac{d \sigma}{\sigma_{inel}\, d y_j} \\
 & \propto & \exp{\left[ - \lambda\, (y_i - y_j) \right] }\ ,
\label{eq34}
\eeqar
and therefore, the particles with large rapidity difference are
uncorrelated. Consider now the inclusive process $1 + 2 \rightarrow
i + X$. Its single particle inclusive cross section
\beq
\ds
\nonumber
f_i  \equiv  E\, \frac{d^3 \sigma_i}{d^3 p} 
  =  \frac{d^2 \sigma (y_1 - y_i, y_i -y_2, p^2_{i\, T})}
{d y_i d^2 p_{i\, T}}
\label{eq35}
\eeq
becomes independent of $y_i - y_2$ at sufficiently high collision
energy in the fragmentation region of particle {\it 1}, provided 
$y_1 - y_i \approx 1$ and $y_i - y_2 \approx y_1 - y_2 \gg 1$. Thus,
the inclusive densities $n_i \equiv f_i / \sigma_{inel}$ are 
determined by only two variables
\beq
\ds
n_i = \phi(y_1 - y_i,\, p^2_{i\, T}) \ .
\label{eq36}
\eeq
Recalling that the Feynman variable $x_F$ is connected to rapidity via
\beq
\ds
x_{i\, F} \equiv \frac{p_{i\, \Vert}}{p_{\Vert}^{max}} 
\approx \exp{ \left[ -(y_1 - y_i) \right] } \ ,
\label{eq37}
\eeq
one arrives from Eq.(\ref{eq22}) to the condition of Feynman scaling
\beq
\ds
n_i = \psi (x^{(i)}_F,\, p^2_{i\, T})\ .
\label{eq38}
\eeq
The invariant distribution 
\beq \ds 
F(x_F) = \frac{2}{\pi \sqrt{s}} \int 
E_{cm}\frac{d^2\sigma}{dx_F dp^2_{\rm T}} dp^2_{\rm T}
\label{eq39}
\eeq
is displayed in Fig.~\ref{fig10} for all charged particles from the
$pp$ collisions at energies from $\sqrt{s} = 200$\,GeV to 14\,TeV. 
The scaling seems to hold within 20\% of accuracy in the fragmentation 
region at $0.1 < x_F < 0.2$ only.

\vspace{0.7cm}

\subsection{Violation of KNO scaling}
\label{KNO}

Another scaling dependence is known as Koba-Nielsen-Olesen or 
KNO scaling \cite{KNO_scal}. Initially it was also derived from the 
hypothesis of Feynman scaling, but later on it appeared that both 
hypotheses are of independent origin. The KNO scaling claims that 
at $\sqrt{s} \rightarrow \infty$ the normalized multiplicity
distribution just scales up as $\ln{s}$ or, equivalently, that
\beq
\ds
\frac{\langle n \rangle \, \sigma_n}{\Sigma\, \sigma_n} = 
\Psi\left( \frac{n}{\langle n \rangle} \right) \ ,
\label{eq40}
\eeq
with $\sigma_n$ being the partial cross section for $n$-particle 
production, $\langle n \rangle$ the average multiplicity and 
$\Psi (n/\langle n \rangle )$ energy independent function. 
KNO scaling was found to hold up to ISR energies, $\sqrt{s} \leq 
62$\,GeV, despite the apparent failure of the Feynman scaling 
hypothesis in the central region $|x_{\rm F} | \leq x_0$. Violation of 
KNO scaling was predicted within the RFT in \cite{AGK,qgsm_1}. Later on 
the violation was observed experimentally by the UA5 and UA1 
Collaborations in $\bar{p}p$ collisions at $\sqrt{s} = 546$\,GeV 
\cite{UA5_rep}. The origin of this phenomenon in the 
model is the following. At ultrarelativistic energies the main 
contribution to particle multiplicity comes from the cut-Pomerons,
and each cut results in formation of two strings. Short range 
correlations inside a single string lead to a Poisson-like 
multiplicity distribution of produced secondaries. At energies below
100\,GeV, the multistring (or chain) processes are not very abundant
and invariant masses of the strings are not very large. Therefore,
different contributions to particle multiplicity overlap strongly, and
KNO scaling is nearly fulfilled. With rising $\sqrt{s}$, the number of
strings increases as $(s/s_0)^\Delta$, and their invariant masses
increase as well. This leads to enhancement of high multiplicities,
deviation of the multiplicity distribution from the Poisson-like
behavior and violation of KNO scaling \cite{mc_qgsm,qgsm_1}. 

\begin{figure}[htb]
 \resizebox{\linewidth}{!}{
\includegraphics{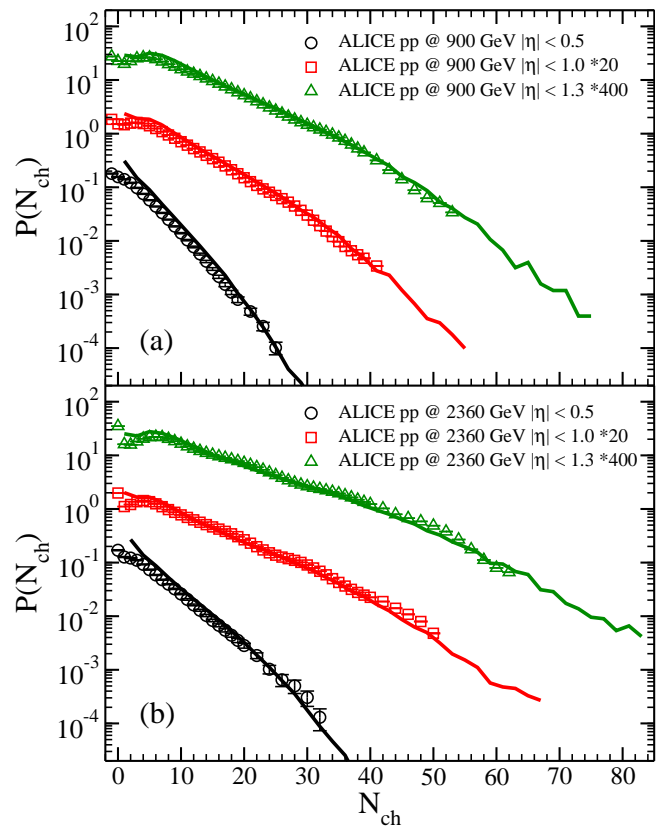}
}
\caption{
(Color online)
Charged particle multiplicity distributions in $|\eta| < 0.5$, 
$|\eta| < 1.0$ and $|\eta| < 1.3$ intervals, obtained in QGSM for
$pp$ collisions at $\sqrt{s} = 900$\,GeV (upper plot) and at
$\sqrt{s} = 2360$\,GeV (bottom plot). Open symbols show the 
corresponding ALICE data \cite{ALICE_2}.
}
\label{fig11}
\end{figure}
Before studying the violation of KNO scaling at LHC, we compare in
Fig.~\ref{fig11} the QGSM calculations with the ALICE data. In this 
figure the multiplicity distributions of charged particles calculated 
in NSD $pp$ events at $\sqrt{s} = 900$\,GeV and $\sqrt{s} = 2.36$
\,TeV in three central pseudorapidity intervals are plotted onto the 
experimental data. The agreement between the model results and the 
data is good. Moreover, the QGSM demonstrates a kind of a wavy 
structure mentioned in \cite{ALICE_2}. As we see below, such a 
wavy behavior in the model can be linked to processes going via the
many-Pomeron exchanges.

\begin{figure}[htb]
 \resizebox{\linewidth}{!}{
\includegraphics{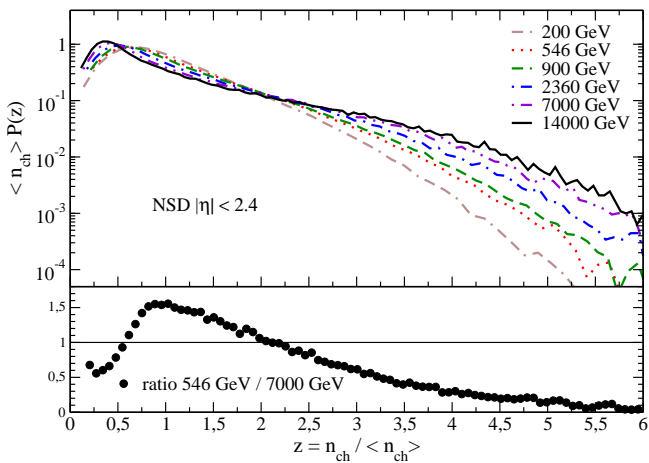}
}
\caption{
(Color online)
Charged particle multiplicity distributions in the KNO variables in
QGSM nondiffractive $pp$ ($p\bar{p}$) collisions at 
$\sqrt{s} =
  200$\,GeV, 564\,GeV, 900\,GeV, 2.36\,TeV, 7\,TeV and 14\,TeV.
}
\label{fig12}
\end{figure}

\begin{figure}[htb]
 \resizebox{\linewidth}{!}{
\includegraphics{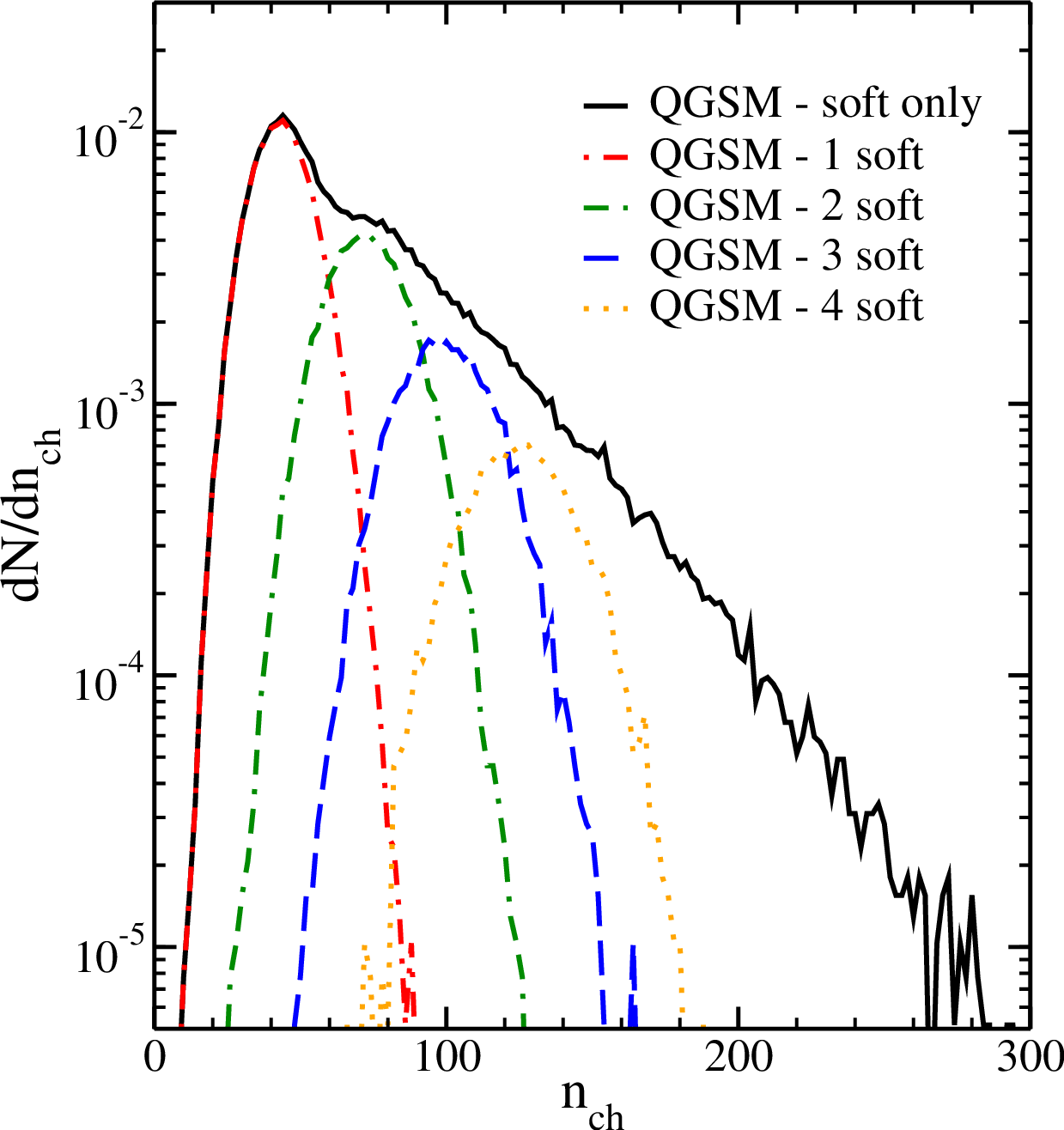}
}
\caption{
(Color online)
Charged particle multiplicity distribution (solid line) for processes 
going via the exchange of $n \geq 1$ soft Pomerons in $pp$ collisions 
at $\sqrt{s} = 14$\,TeV. Contributions of the first four terms are shown 
by dash-dotted ($n=1$), double-dash--dotted ($n=2$), dashed ($n=3$)
and dotted ($n=4$) lines, respectively.
}
\label{fig13}
\end{figure}
The multiplicity distributions of charged particles obtained in QGSM 
for NSD $pp$ collisions at all energies in question are presented in 
Fig.~\ref{fig12} for the interval $|\eta| < 2.4$. Although the 
differences between the neighbor energies seem not to be very dramatic, 
the tendency in the modification of the distributions is quite clear. 
The high-multiplicity tail is pushed up, maximum of the distribution 
is shifted towards small values of $n_{ch}/\langle n_{ch} \rangle$ and 
the characteristic ``shoulder" in the spectrum becomes quite distinct,
as presented by the distribution for top LHC energy. Another 
interesting observation is the unique intersection point for all
distributions. All curves cross each other at $z \approx 2.3$ as can 
be clearly seen in the bottom plot of Fig.~\ref{fig12}, where the
ratio $\langle n_{ch} \rangle P(z)|_{546\, {\rm GeV}} / \langle n_{ch} 
\rangle P(z)|_{7\, {\rm TeV}}$ is displayed. 
This prediction is in line with a recent measurement of multiplicity 
distributions and investigation of KNO scaling \cite{Zaccolo:qm15}, 
where the same qualitative behaviour has now been observed. 
Note that the aforementioned
pseudorapidity range $|\eta| < 2.4$ is not sufficient to observe the 
multihump structure in the KNO plot predicted in \cite{Kaid99} (see
also \cite{KP2}) for the full phase space. 
To clarify the role of multi-Pomeron processes in violation of 
KNO scaling explicitly, Fig.~\ref{fig13} shows the contribution to the 
particle multiplicity diagram coming from the processes with different 
number of soft Pomerons in $pp$ collisions at $\sqrt{s} = 14$\,TeV. 
The maxima of distributions for multi-Pomeron processes are moved in 
the direction of high multiplicities thus lifting the high-multiplicity
tail. The pronounced peak in the low-multiplicity interval arises 
solely due to single Pomeron exchange.   

\section{Conclusions}
\label{sec4}

Multiplicity, transverse momentum and (pseudo)rapidity distributions 
of hadrons produced in $pp$ interactions at energies from $\sqrt{s} =
200$\,GeV to 14\,TeV are studied within the Monte Carlo quark-gluon 
string model. Parameters of soft and hard Pomerons are determined 
from the fit to recent LHC $pp$ data. Compared to the fit to lower
energies, it was found necessary both to increase the intercept of 
soft Pomeron and to reduce its slope parameter. Other parameters, 
such as total cross sections, cross sections of single-diffractive 
and double-diffractive processes, etc, are taken from theoretical 
considerations. The model simulations of pseudorapidity, transverse
momentum and multiplicity spectra of secondaries are in a good 
agreement with the corresponding experimental data obtained in 
$\bar{p}p$ and $pp$ collisions at Tevatron and at CERN energies. 
Predictions are made for $pp$ interactions at top LHC energy 
$\sqrt{s} = 14$\,TeV. 
We demonstrated how an increase of the mean $p_{\rm T}$
with energy is generated within our model due to the interplay of an
increasing exchange of hard Pomerons and explicitly $s$-dependent
$p_{\rm T}$ distributions during the string--break procedure. 
It is shown that within the examined energy range one cannot distinguish 
between the ``standard'' logarithmic dependence ($\propto \ln^2{s}$) 
and novel power-law approximation ($\propto E^\lambda$), employed for 
particle densities and for their mean $p_{\rm T}$ in the present model, 
based on Reggeon Field theory, and in theory of color glass condensate, 
respectively.
  
Several scaling properties 
observed in particle production at relativistic energies have been
examined. QGSM favors violation of Feynman scaling in the central 
rapidity region and its preservation in the fragmentation areas.
Extended longitudinal scaling is shown to hold at LHC. This scaling 
is also attributed to heavy-ion collisions at energies up to
$\sqrt{s} = 200$\,GeV. Extrapolations based on statistical thermal 
model predict its violation at LHC, thus implying vanishing of
Feynman scaling for nuclear collisions in the fragmentation regions
as well. This important problem should be resolved experimentally in
the nearest future. Finally, further violation of the KNO scaling in
multiplicity distributions is demonstrated in QGSM. The origin of 
both conservation and violation of the scaling trends is traced to 
short range correlations of particles in the strings and interplay 
between the multi-Pomeron processes at ultrarelativistic energies.

\begin{acknowledgments}
Fruitful discussions with K.~Boreskov, L.~Csernai, O.~Kancheli,
L.~McLerran and S.~Ostapchenko are gratefully acknowledged.
This work was supported by the Norwegian Research Council (NFR)
under Contract No. 185664/V30. J.B. thanks the German Research 
Foundation (DFG) for the financial support through the Project 
BL~1286/2-1. L.B. acknowledges financial support of the
Alexander von Humboldt Foundation. 
\end{acknowledgments}



\begin{thebibliography}{9}

\bibitem{QM14} {\it Proceedings of Quark Matter 2014\/}, edited by 
P.~Braun-Munzinger, B.~Friman, and J.~Stachel, 
Nucl. Phys. {\bf A931}, 1c (2014).

\bibitem{Abelev:2013sqa}
ALICE Collaboration (B.~Abelev {\it et al.}),
  JHEP {\bf 1309}, 049 (2013). 

\bibitem{CDP08} L.~Cunqueiro, J.~Dias~de~Deus, and C.~Pajares,
Eur. Phys. J. C {\bf 65}, 423 (2010).

\bibitem{Ent10} D.~d'Enterria, G.~Eyyubova, V.~Korotkikh, I.~Lokhtin, 
S.~Petrushanko, L.~Sarycheva, and A.~Snigirev,
Eur. Phys. J. C {\bf 66}, 173 (2010).

\bibitem{BKK08} K.G.~Boreskov, A.B.~Kaidalov, and O.V.~Kancheli,
Eur. Phys. J. C {\bf 58}, 445 (2008).

\bibitem{GLMT16} E.~Gotsman, E.~Levin, U.~Maor, and S.~Tapia,
arXiv:1603.02143 [hep-ph].
  
\bibitem{fritiof}
B.~Andersson, G.~Gustafson, and B.~Nielsson-Almqvist,
Nucl. Phys. {\bf B281}, 289 (1987).

\bibitem{pythia}
H.-U.~Bengtsson and T.~Sjostrand,
Comp. Phys. Commun. {\bf 46}, 43 (1987).
 
\bibitem{mc_qgsm} N.S.~Amelin and L.V.~Bravina,
Sov. J. Nucl. Phys. {\bf 51}, 133 (1990) [Yad. Fiz. {\bf 51}, 211
(1990)];
N.S.~Amelin, L.V.~Bravina, L.I.~Sarycheva, and L.I.~Smirnova,
Sov. J. Nucl. Phys. {\bf 51}, 535 (1990) [Yad. Fiz. {\bf 51}, 841
(1990)].

\bibitem{dpm} 
A.~Capella, U.~Sukhatme, C.-I.~Tan, and J.~Tran Thanh Van,
Phys. Rep. {\bf 236}, 225 (1994).

\bibitem{venus} K.~Werner,
Phys. Rep. {\bf 232}, 87 (1993).

\bibitem{hijing} M.~Gyulassy and X.~N.~Wang, 
Comput. Phys. Commun. {\bf 83}, 307 (1994).

\bibitem{urqmd}
M.~Bleicher,     
E.~Zabrodin, C.~Spieles, S.A.~Bass, C.~Ernst, S.~Soff, L.~Bravina,
M.~Belkacem, H.~Weber, H.~St{\"o}cker, and W.~Greiner,
J. Phys. G {\bf 25}, 1859 (1999).

\bibitem{epos} K.~Werner, F.-M.~Liu, and T.~Pierog,
Phys. Rev. C {\bf 74}, 044902 (2006).

\bibitem{phojet} R.~Engel, J.~Ranft, and S.~Roesler, 
Phys. Rev. D {\bf 52}, 1459 (1995). 

\bibitem{qgsjet2} S.~Ostapchenko,
Nucl. Phys., Proc. Suppl. {\bf 151}, 143 (2006);
Phys. Rev. D {\bf 83}, 014018 (2011).

\bibitem{paciae} Ben-Hao~Sa, Dai-Mei~Zhou, Yu-Liang~Yan, 
Xiao-Mei~Li, Sheng-Qin~Feng, Bao-Guo~Dong, Xu~Cai,
Comput. Phys. Commun. {\bf 183}, 333 (2012).

\bibitem{jetset} 
T.~Sj\"ostrand, Comp. Phys. Commun. {\bf 39}, 347 (1986).

\bibitem{Fer50} E.~Fermi, Prog. Theor. Phys. {\bf 5}, 570 (1950).

\bibitem{Land53} L.D.~Landau,
Izv. Akad. Nauk SSSR, Ser. Fiz. {\bf 17}, 51 (1953) (in Russian);
S.Z.~Belenkij and L.D.~Landau,
Nuovo Cimento, Suppl. {\bf 3}, 15 (1956).

\bibitem{PDG} C.~Amsler {\it et al.\/} (Particle Data Group),
Phys. Lett. B {\bf 667}, 1 (2008).

\bibitem{Feyn_scal} R.~Feynman, 
Phys. Rev. Lett. {\bf 23}, 1415 (1969);
R.~Feynman, 
{\it Photon-Hadron Interactions\/} (Benjamin, New York, 1972).

\bibitem{ext_long_scal} 
G.J.~Alner {\it et al.} (UA5 Collaboration), 
Z. Phys. {\bf C33}, 1 (1986). 

\bibitem{KNO_scal} Z.~Koba, H.B.~Nielsen, and P.~Olesen, 
Nucl. Phys. {\bf B40}, 317 (1972).

\bibitem{Busza08} W.~Busza, J. Phys. G {\bf 35}, 044040 (2008).

\bibitem{CGC} L.~McLerran,
Lect. Notes Phys. {\bf 583}, 291 (2002). 

\bibitem{Lev10} E.~Levin and A.~H.~Rezaeian,
Phys. Rev. D {\bf 82}, 014022 (2010).

\bibitem{MLP10} L.~McLerran and M.~Praszalowicz,
Acta Phys. Pol. B {\bf 41}, 1917 (2010). 

\bibitem{qgsm_1} A.B.~Kaidalov, Phys. Lett. {\bf 116B}, 459 (1982);
A.B.~Kaidalov and K.A.~Ter-Martirosyan, 
Phys. Lett. {\bf 117B}, 247 (1982).


\bibitem{RFT}
V.~Gribov, Sov. Phys. JETP {\bf 26}, 414 (1968);
L.V.~Gribov, E.M.~Levin, and M.G.~Ryskin,
Phys. Rep. {\bf 100}, 1 (1983).

\bibitem{tH74} G.~t'Hooft, Nucl. Phys. {\bf B75}, 461 (1974).

\bibitem{Ven74} G.~Veneziano, Phys. Lett. {\bf 52B}, 220 (1974).

\bibitem{CMV75} M.~Cifaloni, G.~Marchesini, and G.~Veneziano,
Nucl. Phys. {\bf B98}, 472 (1975).

\bibitem{Kaid99} 
A.B.~Kaidalov, Surv. High Energy Phys. {\bf 13}, 265 (1999).

\bibitem{AGK} V.~Abramovskii, V.~Gribov, and O.~Kancheli,
Sov. J. Nucl. Phys. {\bf 18}, 308 (1974) 
[Yad. Fiz. {\bf 18}, 595 (1973)].

\bibitem{BTM76} M.~Baker and K.A.~Ter-Martirosyan,
Phys. Rep. {\bf 28C}, 1 (1976).

\bibitem{KP1} A.B.~Kaidalov and M.G.~Poghosyan, 
arXiv:0909.5156.

\bibitem{KP2} A.B.~Kaidalov and M.G.~Poghosyan, 
Eur. Phys. J. C {\bf 67}, 397 (2010). 

\bibitem{CTVK87}
A.~Capella, J.~Tran~Thanh~Van, and J.~Kweicinski,
Phys. Rev. Lett. {\bf 58}, 2015 (1987).

\bibitem{ASC92} N.S.~Amelin, E.F.~Staubo, and L.P.~Csernai,
Phys. Rev. D {\bf 46}, 4873 (1992).

\bibitem{ICTV88}
V.~Innocente, A.~Capella, and J.~Tran~Thanh~Van,
Phys. Lett. {\bf 213B}, 81 (1988).

\bibitem{Baier_struc} R.~Baier, J.~Engels, and B.~Petersson,
Z. Phys. {\bf C2}, 265 (1979). 

\bibitem{FFF_crosec} R.~P.~Feynman, R.~D.~Field, and G.~C.~Fox
Phys. Rev. D {\bf 18}, 3320 (1978).

\bibitem{enh_diag} O.V.~Kancheli, JETP Lett. {\bf 11}, 267 (1970);
A.H.~Mueller, Phys. Rev. D {\bf 2}, 2963 (1970).

\bibitem{AGK-viol1} Y.~V.~Kovchegov and K.~Tuchin,
Phys. Rev. D {\bf 65}, 074026 (2002).

\bibitem{AGK-viol2} A.~Kovner and M.~Lublinsky,
JHEP {\bf 0611}, 083 (2006).

\bibitem{AGK-viol3} E.~Levin and A.~Prygarin,
Phys. Rev. C {\bf 78}, 065202 (2008).

\bibitem{FF_frag} R.D.~Field and R.P.~Feynman, 
Nucl. Phys. {\bf B136}, 1 (1978).

\bibitem{ber_tue} 
L.V.~Bravina, L.P.~Csernai, P.~Levai, N.S.~Amelin, and D.~Strottman,
Nucl. Phys. {\bf A566}, 461c (1994);
L.V.~Bravina, I.N.~Mishustin, J.P.~Bondorf, Amand ~Faessler, and 
E.E.~Zabrodin, Phys. Rev. C {\bf 60}, 044905 (1999);
E.E.~Zabrodin, C.~Fuchs, L.V.~Bravina and Amand~Faessler,
Phys. Lett. B {\bf 508}, 184 (2001).

\bibitem{ell_fl} G.~Burau, J.~Bleibel, C.~Fuchs, Amand~Faessler, 
L.V.~Bravina, and E.E.~Zabrodin, Phys. Rev. C {\bf 71}, 054905 (2005);
J.~Bleibel, G.~Burau, Amand~Faessler, and C.~Fuchs,
Phys. Rev. C {\bf 76}, 024912 (2007);
J.~Bleibel, G.~Burau, and C.~Fuchs, Phys. Lett. B {\bf 659}, 520 (2008). 

\bibitem{UA5_rep} G.J.~Alner {\it et al.} (UA5 Collaboration), 
Phys. Rep. {\bf 154}, 247 (1987).

\bibitem{UA1} G.~Arnison {\it et al.} (UA1 Collaboration), 
Phys. Lett. {\bf 118B}, 167 (1982);
C.~Albajar {\it et al.} (UA1 Collaboration), 
Nucl. Phys. {\bf B335}, 261 (1990).

\bibitem{CDF} F.~Abe {\it et al.} (CDF Collaboration),
Phys. Rev. Lett. {\bf 61}, 1819 (1988); 
Phys. Rev. D {\bf 41}, R2330 (1990).

\bibitem{E735} T.~Alexopoulos {\it et al.} (E735 Collaboration),
Phys. Rev. D {\bf 48}, 984 (1993).

\bibitem{ALICE_1} ALICE Collaboration (K. Aamodt {\it et al.\/}),
Eur. Phys. J. C {\bf 68}, 89 (2010). 

\bibitem{ALICE_2} ALICE Collaboration (K. Aamodt {\it et al.\/}),
Eur. Phys. J. C {\bf 68}, 345 (2010). 

\bibitem{ALICE_3} ALICE Collaboration (K. Aamodt {\it et al.\/}),
Phys. Lett. B {\bf 693}, 53 (2010). 

\bibitem{ALICE:2013bva} K. Aamodt {\it et al.\/} (ALICE Collaboration), 
Report No. ALICE-PUBLIC-2013-001.

\bibitem{Adam:2015pza}
  ALICE Collaboration (J.~Adam {\it et al.}),
  Phys. Lett. B {\bf 753}, 319 (2016). 

\bibitem{CMS_1} CMS Collaboration (K. Khachatryan {\it et al.\/}),
J. High Energy Phys. 02 (2010) 041. 

\bibitem{CMS_2} CMS Collaboration (K. Khachatryan {\it et al.\/}),
Phys. Rev. Lett. {\bf 105}, 022002 (2010). 

\bibitem{Khachatryan:2015jna}
  CMS Collaboration (V.~Khachatryan {\it et al.}),
  Phys. Lett. B {\bf 751}, 143 (2015). 

\bibitem{Chatrchyan:2014qka} 
  CMS and TOTEM Collaborations (S.~Chatrchyan {\it et al.}),
  Eur. Phys. J. C {\bf 74}, 3053 (2014).

\bibitem{GLMM}
  E.~Gotsman, E.~Levin, U.~Maor, and J.~S.~Miller, 
  Eur. Phys. J. C {\bf 57}, 689 (2008).
 
\bibitem{GLM}
  E.~Gotsman, E.~Levin, and U.~Maor,
  Eur. Phys. J. C {\bf 71}, 1553 (2011).

\bibitem{KMR_1}
  M.~G.~Ryskin, A.~D.~Martin, and V.~A.~Khoze,
  Eur. Phys. J. C {\bf 54}, 199 (2008).

\bibitem{KMR_2}
  M.~G.~Ryskin, A.~D.~Martin, V.~A.~Khoze, and A.~G.~Shuvaev,
  J. Phys. G {\bf 36}, 093001 (2009).

\bibitem{Abelev:2012sea}
  ALICE Collaboration (B.~Abelev {\it et al.}),
  Eur. Phys. J. C {\bf 73}, 2456 (2013). 

\bibitem{Antchev:2013gaa}
  G.~Antchev {\it et al.} (TOTEM Collaboration),
  Europhys. Lett.  {\bf 101}, 21002 (2013).

\bibitem{Aaij:2014vfa}
  R.~Aaij {\it et al.} (LHCb Collaboration),
  J. High Energy Phys. 02 (2015) 129.

\bibitem{CST_08} J.~Cleymans, J.~Str\"umpfer, and L.~Turko,
Phys. Rev. C {\bf 78}, 017901 (2008).

\bibitem{Zaccolo:qm15}
   V.~Zaccolo {\it et al.,} (ALICE Collaboration),
   arXiv:1512.05273; 
   V.~Zaccolo,
   Ph.D. Thesis, The Niels Bohr Institute, University of Copenhagen, 2015
   (unpublished). 

\end{thebibliography}
\end{document}